\documentclass[a4paper,fleqn,usenatbib]{mnras}
\usepackage{newtxtext,newtxmath,bm}

\usepackage{graphicx}    

\newcommand{\mr}{\mathrm}
\newcommand{\bvtheta}{\boldsymbol{\vartheta}}
\newcommand{\bd}{\boldsymbol{d}}
\newcommand{\cL}{\mathcal{L}}
\newcommand{\vDelta}{\bm{\mathit{\Delta}}}

\title[Cross-correlation of the tSZ and the 2MRS]
{Joint analysis of the thermal Sunyaev-Zeldovich effect and 2MASS galaxies: Probing gas physics in the local Universe and beyond}

\author[R. Makiya et al.]{
Ryu Makiya,$^{1,2}$\thanks{E-mail: makiya@mpa-garching.mpg.de}
Shin'ichiro Ando $^{3,2}$
and Eiichiro Komatsu,$^{1,2}$
\\
$^{1}$Max-Planck-Institut f\"ur Astrophysik, Karl-Schwarzschild-Str. 1, 85741 Garching, Germany\\
$^{2}$Kavli Institute for the Physics and Mathematics of the Universe, Todai Institutes for Advanced Study, \\
the University of Tokyo, Kashiwa, Japan 277-8583 (Kavli IPMU, WPI) \\
$^{3}$GRAPPA Institute, University of Amsterdam, 1098XH Amsterdam, Netherlands\\
}

\date{Accepted XXX. Received YYY; in original form ZZZ}

\pubyear{2017}

\begin{document}
\label{firstpage}
\pagerange{\pageref{firstpage}--\pageref{lastpage}}
\maketitle

\begin{abstract}
We present a first joint analysis of the power spectra of the thermal Sunyaev-Zeldovich (tSZ) effect measured by the {\it Planck} and the number density fluctuations of galaxies in the 2MASS redshift survey (2MRS) catalog, including their cross-correlation. 
Combining these measurements with the cosmic microwave background (CMB) data and CMB lensing of {\it Planck} assuming a flat $\Lambda$CDM model, we constrain the mass bias parameter as $B = 1.54 \pm 0.098 (1\sigma)$ [$(1-b) = 0.649 \pm 0.041$, where $(1-b) \equiv B^{-1}$], i.e., the {\it Planck} cluster mass should be $35\%$ lower than the true mass.
The mass bias determined by the 2MRS-tSZ cross-power spectrum alone is consistent with that determined by the tSZ auto-power spectrum alone, suggesting that this large mass bias is not due to obvious systematics in the tSZ data.
We find that the 2MRS-tSZ cross-power spectrum is more sensitive to less massive halos than the tSZ auto-power spectrum and it significantly improves a constraint on the mass dependence of the mass bias.
The redshift dependence is not strongly constrained since the multipole range in which high redshift clusters mainly contribute to the tSZ auto is dominated by the contaminating sources.
We conclude that no strong mass or redshift evolution of the mass bias is needed to explain the data.
\end{abstract}



\section{Introduction}
\label{sec:introduction}
The angular power spectrum of the thermal Sunyaev-Zel'dovich (tSZ) effect (\citealt{sunyaev/zeldovich:1972}) is very sensitive to the amplitude of matter density fluctuation, which is characterized by the cosmological parameters $\Omega_{\rm m}$ and $\sigma_8$ (\citealt{komatsu/kitayama:1999, komatsu/seljak:2002}).

Recently the {\it Planck} satellite provided an all-sky map of the tSZ effect and its angular power spectrum (\citealt{planck2015_sz/etal:2016}).
With this, they put a constraint on the combination of the cosmological parameters, $\sigma_8 \Omega_{\rm m}^{3/8}$, independently from the primordial CMB fluctuation.
However, the derived value of $\sigma_8 \Omega_{\rm m}^{3/8}$ depends on the so-called mass bias parameter $B$,
which is introduced to account for the uncertainty in the cluster mass estimation in the {\it Planck} analysis.
It is defined as the ratio of the {\it Planck} cluster mass and the true mass,
\begin{equation}
  B = M_{\rm 500c, true}/M_{{\rm 500c}, {\it Planck}},
\end{equation}
where $M_{\rm 500c}$ is the mass enclosed by the radius $r_{\rm 500c}$ within which the average mass density is 500 times of the critical density of the Universe.
The mass bias $B$ is related to the more commonly used parameter $(1-b)$ as $B = (1-b)^{-1}$.

Using the gas pressure profile of \cite{arnaud/etal:2010}, which is estimated from a local cluster sample observed by {\it XMM-Newton} assuming the hydrostatic equilibrium (HSE),
\cite{planck2015_sz/etal:2016} reported that $B = 1.25$--$1.67$ is required to reconcile the tSZ power spectrum with the combined constraints from the primordial CMB fluctuation, CMB lensing and baryon acoustic oscillations.
Several authors performed a revised analysis of the tSZ power spectrum and obtained similar results (\citealt{horowitz/seljak:2017, hurier/lacasa:2017, salvati/etal:2018, bolliet/etal:2018}).
From the simulation side, \cite{dolag/komatsu/sunyaev:2016} constructed the tSZ map from their Magneticum Pathfinder Simulations and found that the mass bias of $B = 1.2$ gives a reasonable agreement with the observations.

It has been thought that the mass bias mainly arises from the assumption of the HSE with thermal pressure.
The hydrodynamic simulations showed that the HSE mass underestimates the true mass by 5--20\% due to non-thermal pressure support (e.g., \citealt{kay/etal:2004, rasia/etal:2006, rasia/etal:2012, nagai/vikhlinin/kravtsov:2007, piffaretti/etal:2008, lau/etal:2009, maneghetti/etal:2010}; see also \citealt{shi/komatsu:2014, shi/etal:2015, shi/etal:2016}). 
Other effects such as the calibration error of gas temperature in X-ray observations may also contribute to the mass bias (\citealt{schellenberger/etal:2015}).

There are several surveys that have attempted to calibrate cluster masses by weak gravitational lensing assuming that the lensing mass is unbiased (\citealt{linden/etal:2014, hoekstra/etal:2015, smith/etal:2016, penna-lima/etal:2017, medezinski/etal:2018, sereno/etal:2017}).
The derived mass biases differ among those surveys. 
Some studies found consistent results with the {\it Planck}, but the other studies favored a smaller bias ($B = 1.0$--$1.2$).
Since the cluster samples in those surveys have different masses and redshift ranges, it is still unclear whether the differences between surveys come from systematic uncertainties or the mass and/or redshift evolution of the mass bias (\citealt{andreson:2014, sereno/ettori:2017}).

The cross-correlation technique offers a promising way to address this issue.
Taking the cross-correlation between the tSZ map and other observables whose mass and redshift distribution is different from the tSZ effect provides us tomographic information.
It also enables us to check consistency between data sets.
Another advantage is that the cross-correlation measurements are free from contamination in the tSZ map if the contamination does not correlate with the respective observables.
The uniqueness and effectiveness of the cross-correlation technique are proven by recent studies on the cross-correlation of the tSZ with, e.g., weak lensing shear/convergence field (\citealt{hill/spergel:2014, waerbeke:2014, ma/etal:2015, hojjati/etal:2017, osato/etal:2018}), X-ray clusters (\citealt{hajian/etal:2013}), and galaxy groups from the Sloan Digital Sky Survey (\citealt{vikram/etal:2017}).

In this paper, we present a first joint analysis of the power spectra of the tSZ effect and the number density fluctuations of galaxies in the 2MASS redshift survey (2MRS; \citealt{huchra/etal:2012}), which is a spectroscopic follow-up of Two Micron All Sky Survey (\citealt{2mass}), including their cross-correlation. 
The angular auto-power spectrum of the 2MRS galaxies is dominated by galaxies living inside nearby groups and clusters (\citealt{ando/etal:2018}).
By cross-correlating them with the tSZ map, we can investigate physical properties of hot gas in the nearby halos and their relation with galaxy distributions. 
It may also provide a good constraint on the local universe simulations,  e.g., the Magneticum Pathfinder simulation (\citealt{dolag/komatsu/sunyaev:2016}) and ELUCID (\citealt{wang/etal:2014}).

This paper is organized as follows.
In Section \ref{sec:data} we describe the data sets and measurement of auto- and cross-power spectra.
In Section \ref{sec:model} we outline the model of power spectra.
In Section \ref{sec:interpretation} we perform a joint analysis and obtain constraints on the model parameters and discuss their implications for gas physics.
We summarize our results in Section \ref{sec:summary}.
In Appendix \ref{sec:milca_nilc} we investigate the effect of the tSZ map reconstruction methods.
In Appendix \ref{sec:jk} we describe systematic uncertainties in the covariance matrix estimation.
Throughout the paper, we assume a flat $\Lambda$-CDM cosmology.

\section{Data sets}
\label{sec:data}
\subsection{Construction of the maps}
\subsubsection{Compton-Y}
\label{sec:tszmap}
The tSZ effect is characterized by the Compton-Y parameter, which is defined as (\citealt{sunyaev/zeldovich:1972})
\begin{equation}
y = \int n_{\rm e} \frac{k_{\rm B}T_{\rm e}}{m_{\rm e} c^2} \sigma_{\rm T }\;\mr{d}s,
\end{equation}
where $k_{\rm B}$ is the Boltzmann constant, $m_{\rm e}$ is the electron mass, $c$ is the speed of light, $\sigma_{\rm T}$ is the Thomson scattering cross section, $\mr{d}s$ is the distance along the line of sight, and $n_{\rm e}$ and $T_{\rm e}$ are the electron density and temperature.
The temperature distortion of the cosmic microwave background (CMB) caused by the tSZ effect at a frequency $\nu$ is given by
\begin{equation}
\frac{\Delta T}{T_{\rm CMB}} = g(\nu)y.
\end{equation}
Neglecting relativistic corrections, $g(\nu) = x \coth(x/2)-4$ with $x \equiv h\nu/(k_{\rm B}T_{\rm CMB})$ and $T_{\rm CMB} = 2.725$ K.

For the tSZ data, we use the full sky Compton-Y map provided in the {\it Planck} 2015 public data release \citep{planck2015_sz/etal:2016}.
The {\it Planck} public data include two Compton-Y maps, namely, MILCA and NILC, which use different map reconstruction methods.
We estimate the cross-power spectra between the 2MRS map and both Compton-Y maps and find that they agree with each other within the error (see Appendix \ref{sec:milca_nilc} for details).
In the rest of this paper, we show the results for the NILC map unless otherwise noted. 

The tSZ map is contaminated by the strong thermal dust emission from our Galaxy as well as by emission of the infrared (IR) and radio point sources. To reduce the contamination we apply the Galactic mask, which masks $\sim 40\%$ of the sky, and the point source mask provided by the {\it Planck}.
Combining the Galactic mask and the point source mask, the fraction of the sky available for analysis becomes $f_{\rm sky}^{\rm yy}$ $= 0.494$.

\subsubsection{2MRS}
\label{sec:2mrs_map}
\begin{figure}
 \includegraphics[width=\columnwidth]{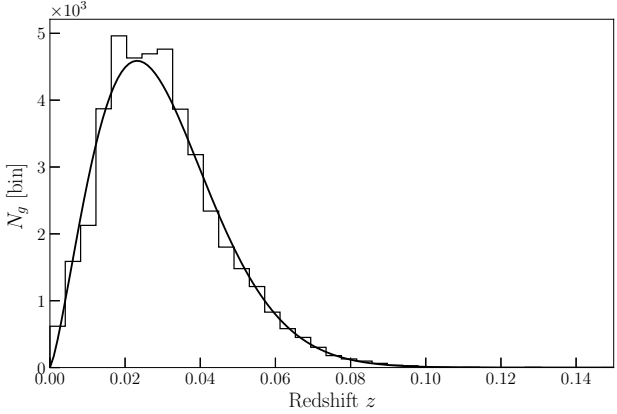}
 \caption{
 The redshift distribution of 2MRS galaxies.
 The black solid line show the fitting function, Eq. (\ref{eq:dndz}).
}.
\label{fig:dndz}
\end{figure}

We use the galaxy catalog obtained by the 2MRS (\citealt{huchra/etal:2012}).
Figure \ref{fig:dndz} shows the redshift distribution of the 2MRS galaxies,
which is well modeled as
\begin{equation}
\label{eq:dndz}
\frac{\mr{d}N_{\rm g}}{\mr{d}z} = N_{\rm g} \frac{ n}{z_0 \Gamma[(m+1)/n]}
\left( \frac{z}{z_0}\right)^{m} \exp \left[-\left(\frac{z}{z_0}\right)^{n}\right],
\end{equation}
with $m = 1.31$, $n = 1.64$, and $z_0 = 0.0266$ \citep{ando/etal:2018}.
The total number of the galaxies in the 2MRS catalogue in the unmasked pixels is $N_{\rm g}$ = 43182.
The median redshift of the 2MRS galaxies is $z_{\rm med} = 0.028$.

From the catalog we construct the pixel-based density fluctuation map of the 2MRS galaxies using the {\tt HEALPix}\footnote{\url{http://healpix.jpl.nasa.gov}} \citep{healpix} package, at the same pixel resolution as the {\it Planck} Compton-Y map, $N_{\rm side} = 2048$. 
The density fluctuation of galaxies $\delta_g$ is defined as
\begin{equation}
\delta_g = \frac{n_g-\bar{n}_g}{\bar{n}_g},
\end{equation}
where $\bar{n}_g = 1.0 \times 10^{-3}$ is the mean number of galaxies per pixel outside the mask, and $n_g$ is the number of galaxies in a given pixel.

Following \cite{ando/etal:2018}, we mask the Galactic plane region and the small regions in which the redshift completeness is low (\citealt{lavaux/hudson:2011}).
With this mask, the sky fraction of the 2MRS galaxy density map is $f_{\rm sky}^{\rm gg} = 0.877$.

For further details of the map construction methods and clustering properties of the 2MRS galaxies, see \cite{ando/etal:2018}.

\subsection{Power spectra}
To compute tSZ auto-, 2MRS auto- and tSZ-2MRS cross-power spectra,
we use the public code {\tt anafast} in the {\tt HEALPix} package (\citealt{healpix}).
The mode coupling effect induced by the mask is corrected by the MASTER algorithm \citep{hivon/etal:2002}.
We use the cross power spectrum of the tSZ and 2MRS masks to correct the masking effect in the tSZ-2MRS cross power spectrum.
The pixelization and beam smearing effects are corrected by dividing the measured power spectra by the square of the window function $w_l$,
\begin{equation}
w_l = 	
\begin{cases}
    p_l b_l & ({\rm tSZ\;auto}) \\[6pt]
    p_l & ({\rm 2MRS\;auto}) \\[6pt]
    \sqrt{p^2_l b_l} & ({\rm tSZ\mathchar`-2MRS\;cross}) \\[6pt]
 \end{cases},
\end{equation}
where $p_l$ is the pixel window function for $N_{\rm side} = 2048$ and $b_l$ is the beam window function of the {\it Planck}.
We assume that the {\it Planck} beam is approximated by a circular Gaussian with the FWHM of $10\arcmin$ \citep{planck2015_sz/etal:2016}.

In the following, we denote an angular power spectrum between observables $A$ and $B$ as $C_{l}^{\rm AB}$. For example, the galaxy auto power spectrum is written as $C_{l}^{\rm gg}$, while the galaxy-tSZ cross power spectrum is $C_{l}^{\rm gy}$.
We bin all the power spectra in 19 bins that are logarithmically equally spaced in multipole $l$ (covering $9 < l < 1411$), weighted by $l(l+1)$.

The measured power spectra are summarized in Table \ref{tb:cl}.

\subsubsection{The tSZ auto power spectrum}
Following \cite{planck2015_sz/etal:2016}, we cross-correlate Compton-Y maps from the first and second halves of the data to obtain the angular auto power spectrum of the tSZ signal.

While the cross-correlation between the first and second halves of the data reduces the instrumental noise bias, there still remains some contamination from the cosmic infrared background (CIB) and residual IR and radio point sources.
In the analysis of the power spectra, we take into account the contamination from those sources (see Section \ref{sec:model_yy} and \ref{sec:mcmc} for details).
 
 In the middle panel of Figure \ref{fig:cl}, we show the measured tSZ auto power spectrum with the contaminating sources subtracted as described in Section \ref{sec:model_yy}.
 
\subsubsection{The 2MRS auto power spectrum}
The galaxy auto power spectrum is measured from the density fluctuation map described in Section \ref{sec:2mrs_map} by also using {\tt anafast} and MASTER.
We employ the same procedure as in \cite{ando/etal:2018} to subtract the shot noise term.
First, we randomly divide the galaxy catalog into two subsets, both of which contain roughly the same number of galaxies.
Then we convert these two subsets into the density fluctuation map, $\delta_{g,1}$ and $\delta_{g,2}$, and construct a half-sum (HS) and a half-difference (HD) map from them:
\begin{equation}
HS = \frac{\delta_{g,1}+\delta_{g,2}}{2},\;\;HD = \frac{\delta_{g,1}-\delta_{g,2}}{2}.
\end{equation}
Since the power spectrum measured from the HS map contains both signal and noise while that from the HD map contains only noise, the galaxy auto power spectrum is estimated as
\begin{equation}
C_{l}^{\rm gg} = C_{l}^{\rm gg, HS}-C_{l}^{\rm gg, HD}.
\end{equation}

In the top panel of Figure \ref{fig:cl} we show the measured galaxy auto power spectrum.

\subsubsection{The tSZ-2MRS cross power spectrum}
\label{sec:cross}
We use {\tt anafast} and MASTER to measure the tSZ-galaxy cross power spectrum from the {\it Planck} full mission Compton-Y map and the HS map of the 2MRS galaxies.

\begin{figure}
 \includegraphics[width=\columnwidth]{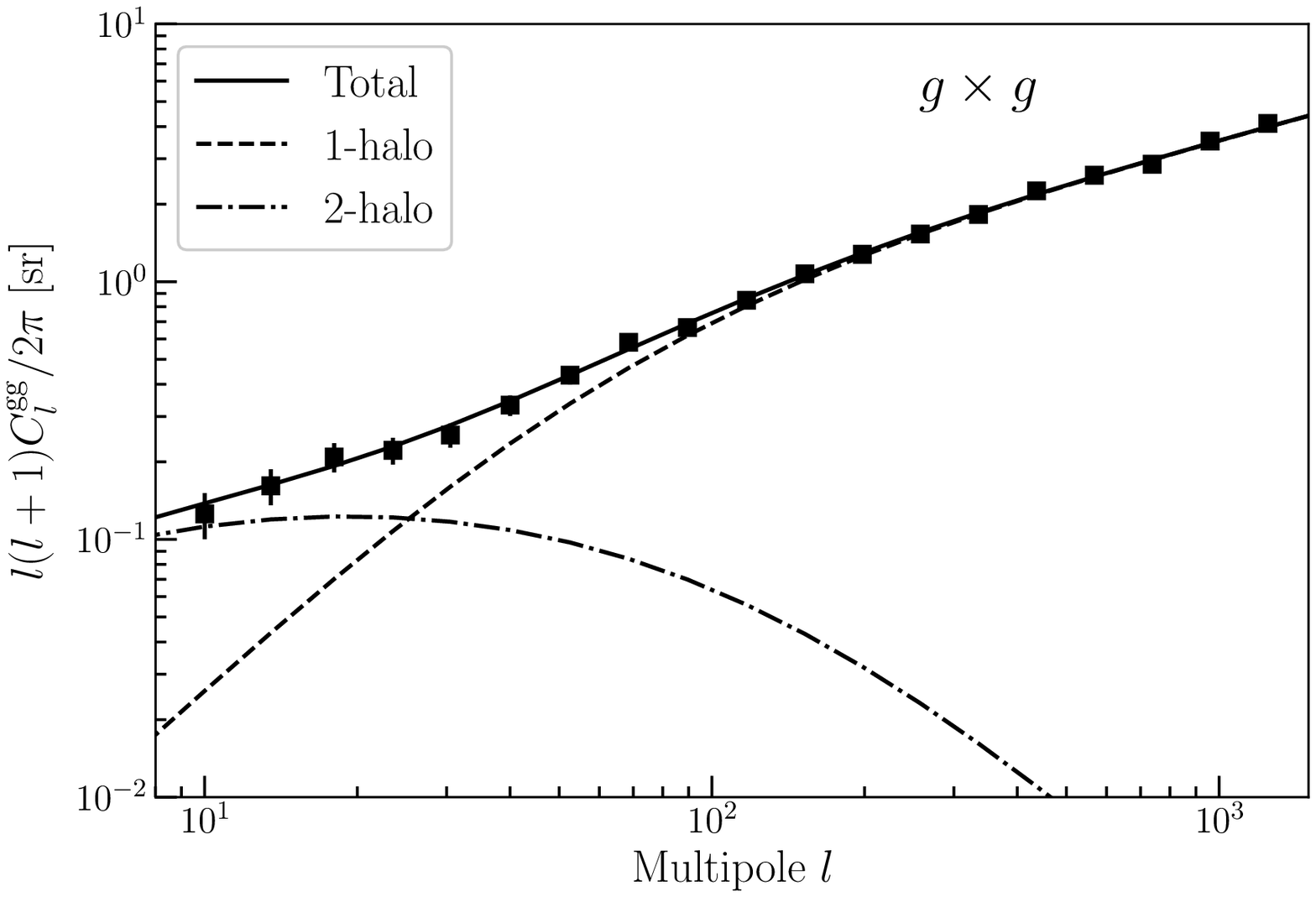}
 \includegraphics[width=\columnwidth]{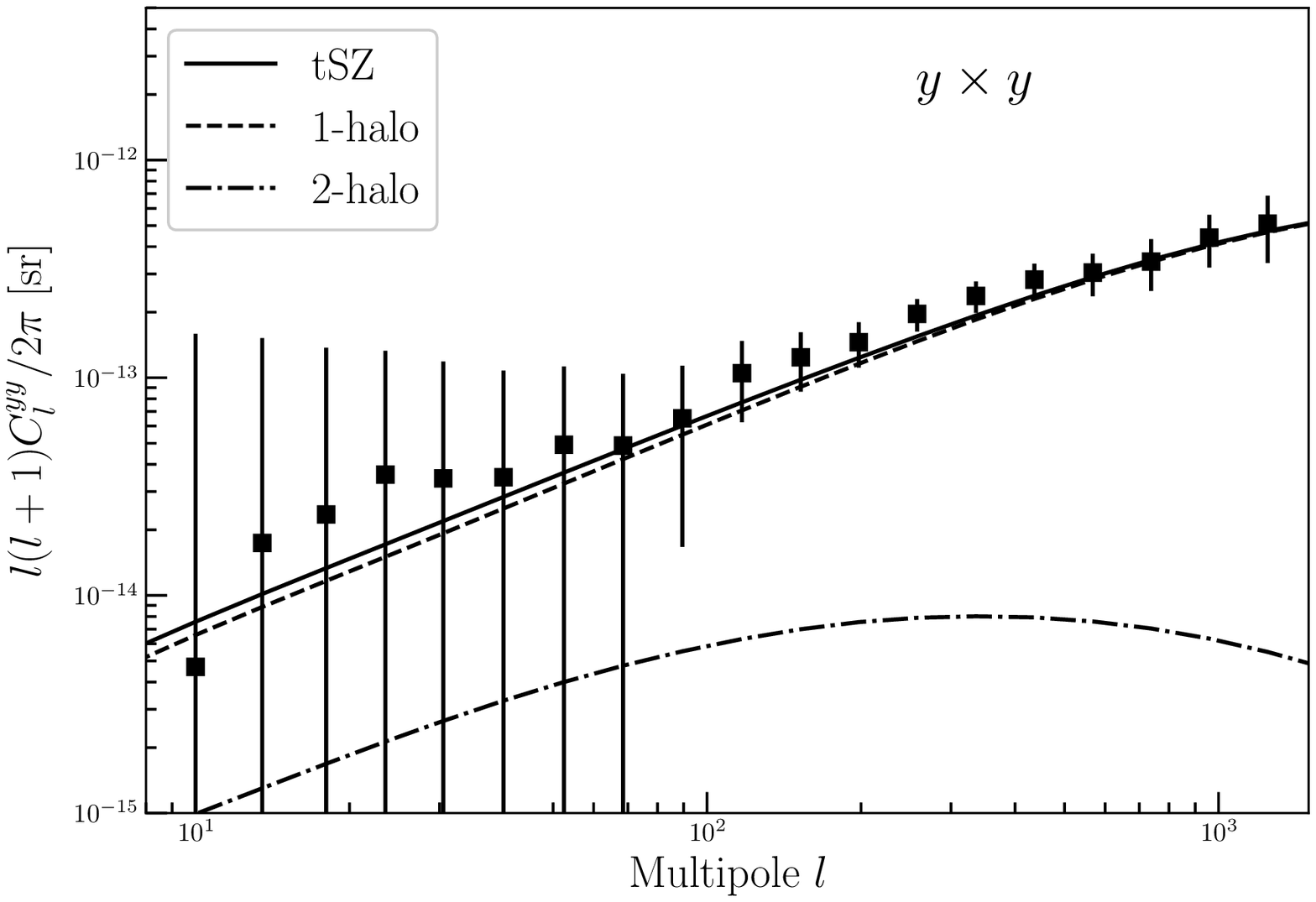}
 \includegraphics[width=\columnwidth]{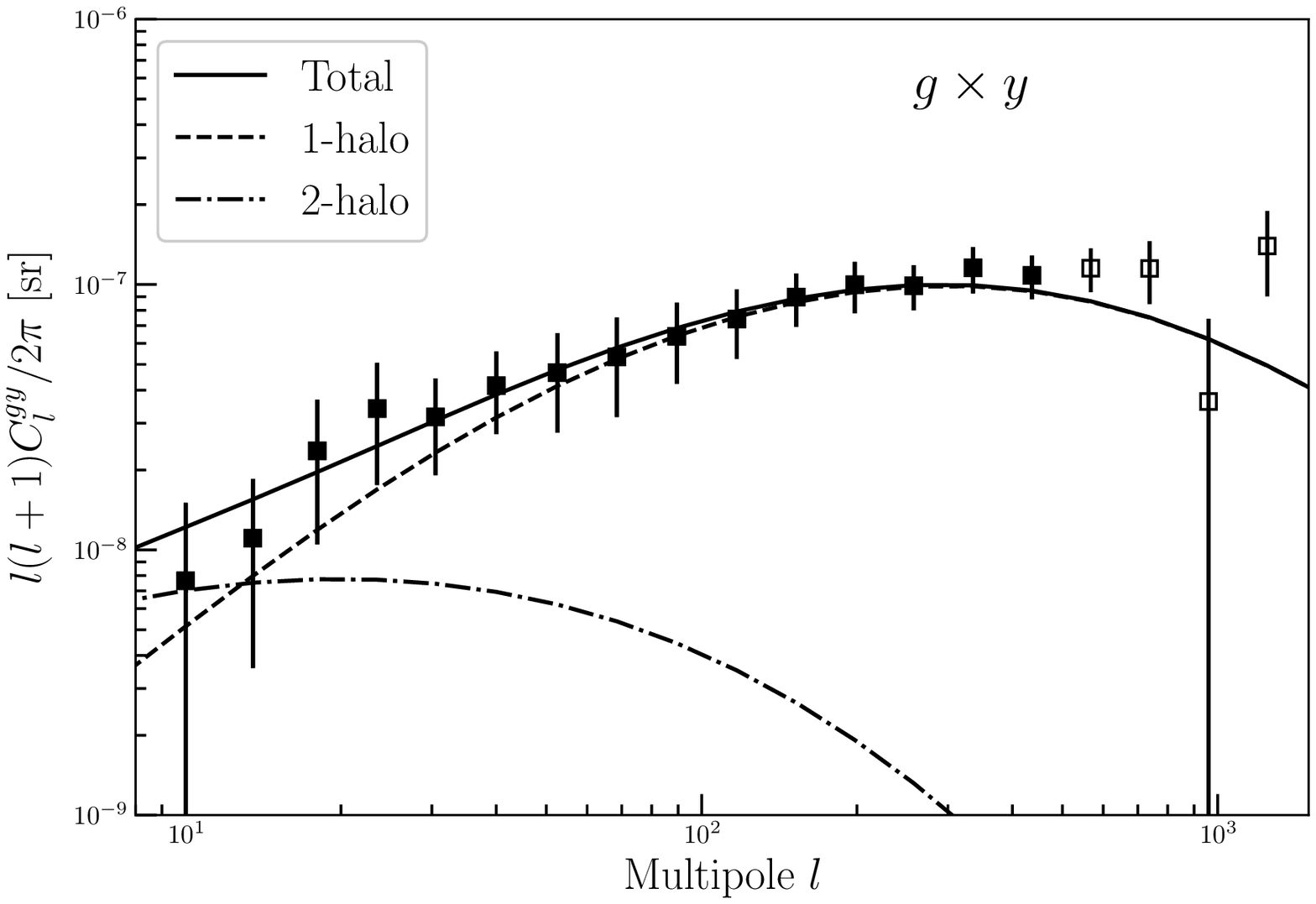}
 \caption{(Top) 2MRS auto-, (Middle) tSZ auto- and (Bottom) 2MRS-tSZ cross-power spectra. The squares show the measured power spectra while the black lines are the best-fitting model described in Section \ref{sec:halomodel}. 
The cross power spectrum data at $l > 500$ shown by the open squares are not used in the parameter fitting (see Section \ref{sec:cross}). 
The dashed lines are the 1-halo (intra-halo) terms, the dot-dashed lines are the 2-halo (inter-halo) terms, and the solid lines are the sum of them. 
The error bars denote the diagonal terms of the covariance matrix which is described in Section \ref{sec:cov}.
For the tSZ power spectrum the error also includes uncertainties in the model of contaminating sources (see Section \ref{sec:model_yy}).
}.
\label{fig:cl}
\end{figure}

In the bottom panel of Figure \ref{fig:cl} we show the measured tSZ-galaxy cross power spectrum.
The signal is detected with the signal-to-noise ratio of 9.92 ($l < 500$), which is calculated as
\begin{equation}
\left( \frac{S}{N} \right)^2 = (C^{\rm gy}_{l=10}, \ldots, C^{\rm gy}_{l=436}) {\rm Cov}^{-1}(C^{\rm gy}_l,C^{\rm gy}_{l'}) (C^{\rm gy}_{l=10}, \ldots, C^{\rm gy}_{l=436})^{\intercal},
\end{equation}
where ${\rm Cov}(C^{\rm gy}_l,C^{\rm gy}_{l'})$ is the covariance matrix of the cross-power spectrum described in Section \ref{sec:cov}.

As shown in \cite{planck2015_sz/etal:2016}, contaminations from the CIB and IR and radio point sources begin to dominate the angular power spectrum of Compton-Y at higher multipoles.
As the CIB is thought to mainly come from relatively higher redshift ($z\sim1$), it should not correlate with the 2MRS galaxies in the local universe ($z < 0.1$).
On the other hand, the redshift distribution of the residuals of IR and radio point sources are still unclear and it could correlate with the 2MRS sources.
Thus we decide not to use $C_l^{\rm gy}$ at $l > 500$ in the parameter fitting.

\begin{table*}
\begin{center}
\begin{tabular}{cccccccccc}
\hline
 $l$ & $C_l^{\rm gg}$ & $\sigma_{\rm G}(C_l^{\rm gg})$ & $\sigma_{\rm NG}(C_l^{\rm gg})$ & $C_l^{\rm yy}$ & $\sigma_{\rm G}(C_l^{\rm yy})$ & $\sigma_{\rm NG}(C_l^{\rm yy})$ & $C_l^{\rm gy}$ & $\sigma_{\rm G}(C_l^{\rm gy})$ & $\sigma_{\rm NG}(C_l^{\rm gy})$ \\
\hline
  10 & 7.18e-03 & 1.42e-03 & 3.28e-04 & 2.68e-16 & 1.52e-16 & 8.83e-15 & 4.37e-10 & 2.31e-10 & 6.06e-10 \\
  13 & 5.18e-03 & 7.77e-04 & 2.65e-04 & 5.59e-16 & 1.47e-16 & 4.32e-15 & 3.55e-10 & 2.08e-10 & 4.03e-10 \\
  18 & 3.84e-03 & 4.52e-04 & 2.09e-04 & 4.32e-16 & 1.13e-16 & 2.08e-15 & 4.34e-10 & 1.27e-10 & 2.69e-10 \\
  23 & 2.42e-03 & 2.39e-04 & 1.61e-04 & 3.92e-16 & 8.12e-17 & 1.06e-15 & 3.72e-10 & 7.94e-11 & 1.81e-10 \\
  30 & 1.66e-03 & 1.31e-04 & 1.21e-04 & 2.25e-16 & 3.63e-17 & 5.49e-16 & 2.07e-10 & 3.87e-11 & 1.21e-10 \\
  40 & 1.27e-03 & 7.74e-05 & 8.63e-05 & 1.34e-16 & 1.76e-17 & 2.79e-16 & 1.59e-10 & 2.02e-11 & 7.82e-11 \\
  52 & 9.71e-04 & 4.85e-05 & 5.96e-05 & 1.10e-16 & 9.90e-18 & 1.41e-16 & 1.04e-10 & 1.26e-11 & 4.90e-11 \\
  68 & 7.68e-04 & 3.13e-05 & 4.05e-05 & 6.44e-17 & 5.00e-18 & 7.28e-17 & 7.04e-11 & 6.92e-12 & 3.02e-11 \\
  89 & 5.15e-04 & 1.76e-05 & 2.68e-05 & 5.05e-17 & 3.10e-18 & 3.74e-17 & 4.95e-11 & 4.05e-12 & 1.79e-11 \\
 117 & 3.86e-04 & 1.14e-05 & 1.73e-05 & 4.78e-17 & 2.17e-18 & 1.92e-17 & 3.38e-11 & 2.77e-12 & 1.02e-11 \\
 152 & 2.88e-04 & 7.39e-06 & 1.10e-05 & 3.33e-17 & 1.38e-18 & 9.89e-18 & 2.41e-11 & 1.70e-12 & 5.64e-12 \\
 198 & 2.04e-04 & 4.90e-06 & 6.86e-06 & 2.32e-17 & 8.45e-19 & 5.16e-18 & 1.59e-11 & 1.05e-12 & 3.00e-12 \\
 257 & 1.45e-04 & 3.26e-06 & 4.19e-06 & 1.85e-17 & 5.73e-19 & 2.68e-18 & 9.35e-12 & 6.91e-13 & 1.52e-12 \\
 335 & 1.02e-04 & 2.23e-06 & 2.52e-06 & 1.32e-17 & 4.03e-19 & 1.38e-18 & 6.43e-12 & 4.52e-13 & 7.23e-13 \\
 436 & 7.41e-05 & 1.60e-06 & 1.50e-06 & 9.29e-18 & 2.80e-19 & 7.17e-19 & 3.56e-12 & 3.07e-13 & 3.28e-13 \\
 567 & 5.04e-05 & 1.15e-06 & 8.99e-07 & 5.92e-18 & 2.09e-19 & 3.71e-19 & 2.24e-12 & 2.05e-13 & 1.41e-13 \\
 738 & 3.29e-05 & 8.37e-07 & 5.41e-07 & 3.94e-18 & 1.65e-19 & 1.91e-19 & 1.32e-12 & 1.42e-13 & 5.73e-14 \\
 959 & 2.40e-05 & 6.27e-07 & 3.30e-07 & 3.00e-18 & 1.46e-19 & 9.78e-20 & 2.47e-13 & 1.02e-13 & 2.21e-14 \\
1247 & 1.66e-05 & 4.76e-07 & 2.05e-07 & 2.06e-18 & 1.58e-19 & 4.95e-20 & 5.64e-13 & 7.57e-14 & 8.14e-15 \\

\hline
\end{tabular}
\caption{The 2MRS auto-, Compton-Y auto- and 2MRS and Compton-Y cross-power spectra. 
The best-fitting models of the contaminating sources are subtracted from the tSZ auto.
The Gaussian and non-Gaussian errors, $\sigma_{\rm G}$ and $\sigma_{\rm NG}$, are also shown. See section \ref{sec:interpretation} for details. }
\label{tb:cl}
\end{center}
\end{table*}

\section{MODEL}
\label{sec:model}
In this section we describe our model of the auto- and cross-power spectra.

We define the virial mass of a dark matter halo ($M \equiv M_{\rm vir}$) as the mass enclosed within the virial radius $r_{\rm vir}$, which is the one that contains average mass density of $\Delta_{\rm vir}(z)$ times the critical density of the Universe at $z = 0$, where $\Delta_{\rm vir} \equiv 18\pi^2+82d-39d^2$ with $d \equiv \Omega_{m}(1+z)^3/[\Omega_m(1+z)^3+\Omega_{\Lambda}]-1$ \citep{bryan/norman:1998}. 
Other definitions of the mass and radius are expressed in the same way, e.g., $r_{200c}$ and $M_{200c}$ denote the radius which encloses the 200 times the critical density and the mass enclosed within that radius.
To convert mass from one definition to another, we use the fitting formulae of \citet{hu/kravtsov:2003} assuming the Navarro-Frenk-White (NFW) density profile \citep{nfw} and the mass-concentration relation of \cite{sanchez-conde/prada:2014}.

\cite{bolliet/etal:2018} reported that the uncertainty in the model of the mass-concentration relation has non-negligible effects on the computed tSZ power spectrum through the mass conversion.
To avoid this uncertainty, we use the mass function for $M_{\rm 500c}$, since the electron pressure profile is parameterized as a function of $M_{\rm 500c}$ and $r_{\rm 500c}$ (see Section \ref{sec:model_yy}).
On the other hand, the galaxy power spectrum is modeled as a function of the virial mass, thus the model parameters related to the galaxy power spectrum depend on the model of the mass-concentration relation.

\subsection{Halo model}
\label{sec:halomodel}
To compute the angular power spectrum, we use the halo model \citep{komatsu/kitayama:1999, seljak:2000}. In this framework the power spectrum is divided into intra-halo (1-halo)  and inter-halo (2-halo) terms as $C_{l}^{\rm AB} = C_{l}^{\rm AB, 1h}+C_{l}^{\rm AB,2h}$. The 1-halo term is defined as
\begin{equation}
\label{eq:1halo}
C_{l}^{\rm AB,1h} = \int_{z_{\rm min}}^{z_{\rm max}} \mr{d}z\frac{\mr{d}V}{\mr{d}z \mr{d}\Omega} 
\int_{M_{\rm min}}^{M_{\rm max}} \mr{d}M \frac{\mr{d}n}{\mr{d}M}
\tilde{u}_{l}^{\rm A}(M,z)\tilde{u}_{l}^{\rm B}(M,z),
\end{equation}
where $\tilde{u}_{l}^{\rm A}(M,z)$ and $\tilde{u}_{l}^{\rm B}(M,z)$ are the 2D Fourier transform of the projected distribution of observables A and B, respectively.
For the model of the dark matter halo mass function, $\mr{d}n/\mr{d}M$, we use the Magneticum Pathfinder simulation \citep{bocquet/etal:2016}, with the parameters for ``$M_{\rm500c}$ Hydro'' which is for $M_{\rm 500c}$ and the baryonic effects are taken into account.
For the mass and redshift range of integration, we find that $1 \times 10^{10} h^{-1} M_{\odot} < M_{\rm 500c} < 1 \times 10^{16} h^{-1} M_{\odot}$ and $1 \times 10^{-5} < z < 6$ suffice to get the integral to converge (see Section \ref{sec:dzdm} and Figure \ref{fig:cldz_cldm} later).

The 2-halo term is written as
\begin{equation}
\label{eq:2h}
C_{l}^{\rm AB,2h} = \int_{z_{\rm min}}^{z_{\rm max}} \mr{d}z\frac{\mr{d}V}{\mr{d}z \mr{d}\Omega} b^{\rm A}_{l}(z)b^{\rm B}_{l}(z)P_\mr{lin}(l/\chi,z),
\end{equation}
where $P_{\rm lin}(k,z)$ is the linear matter power spectrum computed with {\tt CAMB} \citep{lewis/etal:1999,howlett/etal:2012}, and $b_{l}^{\rm A}$ and $b_{l}^{\rm B}$ are the scale dependent bias of the observables A and B, which will be described in Section \ref{sec:model_yy} and \ref{sec:model_gg}.

\subsubsection{tSZ model}
\label{sec:model_yy}
The tSZ term in Eq. (\ref{eq:1halo}), $\tilde{u}_l^{y}$, is given by \citep{komatsu/seljak:2002}
\begin{equation}
\label{eq:y}
\tilde{u}_l^{y}(M,z) = \frac{4\pi r_{\rm 500c}}{l_{\rm 500c}^2}
\int^{x_{\rm max}}_{x_{\rm min}}\mr{d}x x^2 y_\mr{3D}(x)
\frac{\sin(lx/l_\mr{500c})}{lx/l_\mr{500c}},
\end{equation}
where $x \equiv r/r_{\rm 500c}$, $l_{500c} \equiv D_{A}/r_{\rm 500c}$, and $D_{A}$ is the proper angular diameter distance.
The integral is performed between $x_{\rm min} = 1 \times 10^{-6}$ and $x_{\rm max} = 6$.
The radial distribution of Compton-Y, $y_{\rm 3D}(x)$, is written by an electron pressure profile $P_e(x)$ as
\begin{equation}
y_{\rm 3D}(x) = \frac{\sigma_{T}}{m_e c^2}P_e(x).
\end{equation}
We use \citep{arnaud/etal:2010}
\begin{equation}
\begin{split}
\label{eq:pe}
P_e(x) &= 1.65(h/0.7)^2 {\rm \;eV\;cm^{-3}} \\
& \times E^{8/3}(z) \left[ \frac{M_{\rm 500c}}{3 \times 10^{14}(0.7/h)M_{\odot}}\right]^{2/3+\alpha_p}p(x),
\end{split}
\end{equation}
with $E(z) \equiv H(z)/H_0$. 
The generalized NFW profile, $p(x)$, is defined by \citep{nagai/etal:2007}
\begin{equation}
p(x) \equiv \frac{P_{0}(0.7/h)^{3/2}}{(c_{500}x)^{\gamma}
[1+(c_{500}x)^{\alpha}]^{(\beta-\gamma)/\alpha}}.
\end{equation}
We use the best-fitting parameter values determined by the analysis of stacked pressure profiles of {\it Planck} tSZ clusters: $P_{0} = 6.41, $$c_{500} = 1.81$, $\alpha = 1.33$, $\beta = 4.13$, and $\gamma = 0.31$ \citep{planck_inter_v:2013}.

The parameter $\alpha_p$ represents a deviation from the standard self-similar solution, i.e. $\alpha_p = 0$ for self-similar.
\cite{arnaud/etal:2010} find $\alpha_p = 0.12$ from their X-ray sample.

As already mentioned in Section \ref{sec:introduction},
the mass-pressure relation Eq.(\ref{eq:pe}) is calibrated against the X-ray cluster sample whose mass tends to be biased low.
To take into account this effect we introduce the mass bias parameter $B$ and rescale the $M_{\rm 500c}$ and $r_{\rm 500c}$ in Equations (\ref{eq:y}) and (\ref{eq:pe}) to $M_{\rm 500c}/B$ and $r_{\rm 500c}/B^{1/3}$, respectively.
We also consider a redshift evolution of $B$ by introducing another free parameter $\rho$ as $B(z) = B (1+z)^{\rho}$.

For the 2-halo term of the power spectra, the tSZ bias $b_l^{y}$ is written as \citep{komatsu/kitayama:1999}
\begin{equation}
b_l^{y}(z) = \int_{M_{\rm min}}^{M_{\rm max}} \mr{d}M \frac{\mr{d}n}{\mr{d}M} \tilde{u}_l^{y}(M,z) b_{\rm lin}(M,z),
\end{equation}
where $b_{\rm lin}(M,z)$ is the linear halo bias of \cite{tinker/etal:2010}.

In addition to the tSZ term, the auto-power spectrum of the Compton-Y map also contains the signal from the residual foreground sources: clustered CIB; radio point sources; IR point sources; Galactic thermal dust emission; and correlated noise. 
In the multipole range we used in the analysis (9 < $l$ < 1411), we can neglect the residual signals from Galactic thermal dust emission \citep{planck2015_sz/etal:2016}. 
Assuming that the signals from these components do not correlate with each other, the measured Compton-Y map auto power spectrum is modeled as
\begin{equation}
C_{l}^{{\rm yy}, {\rm tot}} = C_{l}^{\rm yy}+A_{\rm CIB}C_{l}^{\rm CIB}+A_{\rm IR}C_{l}^{\rm IR}+A_{\rm Rad}C_{l}^{\rm Rad}+A_{\rm CN}C_{l}^{\rm CN},
\end{equation}
where $C_{l}^{\rm yy}$ is the tSZ power spectrum described in Section \ref{sec:model}, and $C_{l}^{\rm CIB}$, $C_{l}^{\rm IR}$, $C_{l}^{\rm RS}$, and $C_{l}^{\rm CN}$ are the power spectra of the clustered CIB, IR and radio point sources and the correlated noise, respectively. 
For these noise terms, we use the templates taken from \cite{planck2015_sz/etal:2016}.
The normalization factors $A_{\rm CIB}$, $A_{\rm IR}$ and $A_{\rm RS}$ are treated as free parameters.
Following \cite{bolliet/etal:2018} we fix $A_{\rm CN} = 0.903$ to reproduce the $C_{l}^{\rm yy, tot}$ at $l=2742$, since the power spectrum is dominated by the correlated noise at this highest multipole.

\subsubsection{2MRS model}
\label{sec:model_gg}
The galaxy term of Eq. (\ref{eq:1halo}), $\tilde{u}^{g}_l$, is given by (\citealt{ando/etal:2018})
\begin{equation}
\begin{split}
\tilde{u}_{l}^{g} &= \frac{W^g(z)}{\chi^2}\frac{1}{\langle n_g(z) \rangle}\\
&\times \sqrt{2\langle N_{\rm sat}|M_{\rm vir}\rangle \tilde{u}_\mr{sat}(k,M_{\rm vir})
 +\langle N_\mr{sat}|M_{\rm vir} \rangle^2 \tilde{u}_\mr{sat}(k,M_{\rm vir})^2},
\end{split}
\end{equation}
where $\chi$ is the comoving distance to a galaxy at a redshift $z$, 
$W^g(z)$ is the redshift kernel of galaxies defined as $W^g \equiv (\mr{d}n_g/\mr{d}z)(\mr{d}z/\mr{d}\chi)$ and $\mr{d}n_{\rm g}/\mr{d}z$ is the redshift distribution of the 2MRS galaxies defined by Eq. (\ref{eq:dndz}).

The function $\tilde{u}_{\rm sat}(k,M)$ is the Fourier transform of the radial distribution of satellite galaxies, normalized to 1 at $k = 0$. 
We assume that the distribution of satellite galaxies, $u_{\rm sat}$,  follows a truncated NFW profile (\citealt{nfw}), characterized by the scale radius of galaxies $r_{\rm s,g}$ and the maximum radius $r_{\rm max,g}$ as,
\begin{equation}
u_{\rm sat}(r,M) \propto \frac{1}{(r/r_{\rm s,g})(r/r_{\rm s,g}+1)^2}
\Theta(r_{\rm max,g}-r),
\end{equation}
where $\Theta$ is the Heaviside step function.
We assume that $r_{\rm s,g}$ is proportional to the NFW scale radius $r_{\rm s}$, and treat $r_{\rm s,g}/r_{\rm s}$ as a free parameter. To compute $r_{\rm s}$ we again use the mass-concentration relation of \cite{sanchez-conde/prada:2014}.

For the halo occupation distribution (HOD) functions, which determine the mean number of central galaxies, $\langle N_{\rm cen}|M_{\rm vir}\rangle$, and that of satellite galaxies, $\langle N_{\rm sat}|M_{\rm vir}\rangle$, as a function of dark matter halo mass $M_{\rm vir}$, we adopt the following expressions (\citealt{zheng/etal:2005})
\begin{equation}
\langle N_{\rm cen}|M_{\rm vir}\rangle = \frac{1}{2}
\left[1+\mr{erf}\left(\frac{\log{M_{\rm vir}}-\log{M_{\rm min}}}{\sigma_{\log{M}}}\right) \right],
\end{equation}
\begin{equation}
\langle N_{\rm sat}|M_{\rm vir}\rangle = \left( \frac{M_{\rm vir}-M_0}{M_1}\right)^{\alpha_{\rm g}}
\Theta(M_{\rm vir}-M_0),
\end{equation}
where erf is the error function, and $M_{\rm min}$, $\sigma_{\log{M}}$, $M_{0}$, $M_{1}$ and $\alpha_{\rm g}$ are free parameters.
Following \cite{ando/etal:2018}, we fix $\sigma_{\log{M}} = 0.15$ and $M_{0} = M_{\rm min}$.
The latter condition means that satellite galaxies form only in halos that host a central galaxy.
The mean number density of galaxies at a redshift $z$, $\langle n_g(z) \rangle$, is written by the HOD functions as
\begin{equation}
\langle n_g(z) \rangle = \int_{M_{\rm min}}^{M_{\rm max}} \mr{d}M \frac{\mr{d}n}{\mr{d}M}
[\langle N_{\rm cen}|M_{\rm vir}\rangle+\langle N_{\rm sat}|M_{\rm vir}\rangle].
\end{equation}

For the 2-halo term, the scale-dependent galaxy bias $b^{\rm g}_l(z)$ is given by
\begin{equation}
\begin{split}
b^{\rm g}_{l}(z) &= \frac{W^{g}(z)}{\chi^2}
\frac{1}{\langle n_g(z) \rangle}
\int_{M_{\rm min}}^{M_{\rm max}} \mr{d}M \frac{\mr{d}n}{\mr{d}M}
[\langle N_{\rm cen}|M_{\rm vir}\rangle \\
&\quad +\langle N_\mr{sat}|M_{\rm vir} \rangle  \tilde{u}_\mr{sat}(l/\chi,M_{\rm vir})]
b_{\rm lin}(M,z),
\end{split}
\end{equation}
where $b_l (M,z)$ is again the linear halo bias of \cite{tinker/etal:2010}.

\subsection{Redshift and mass distribution}
\label{sec:dzdm}
\begin{figure*}
 \includegraphics[width=\columnwidth]{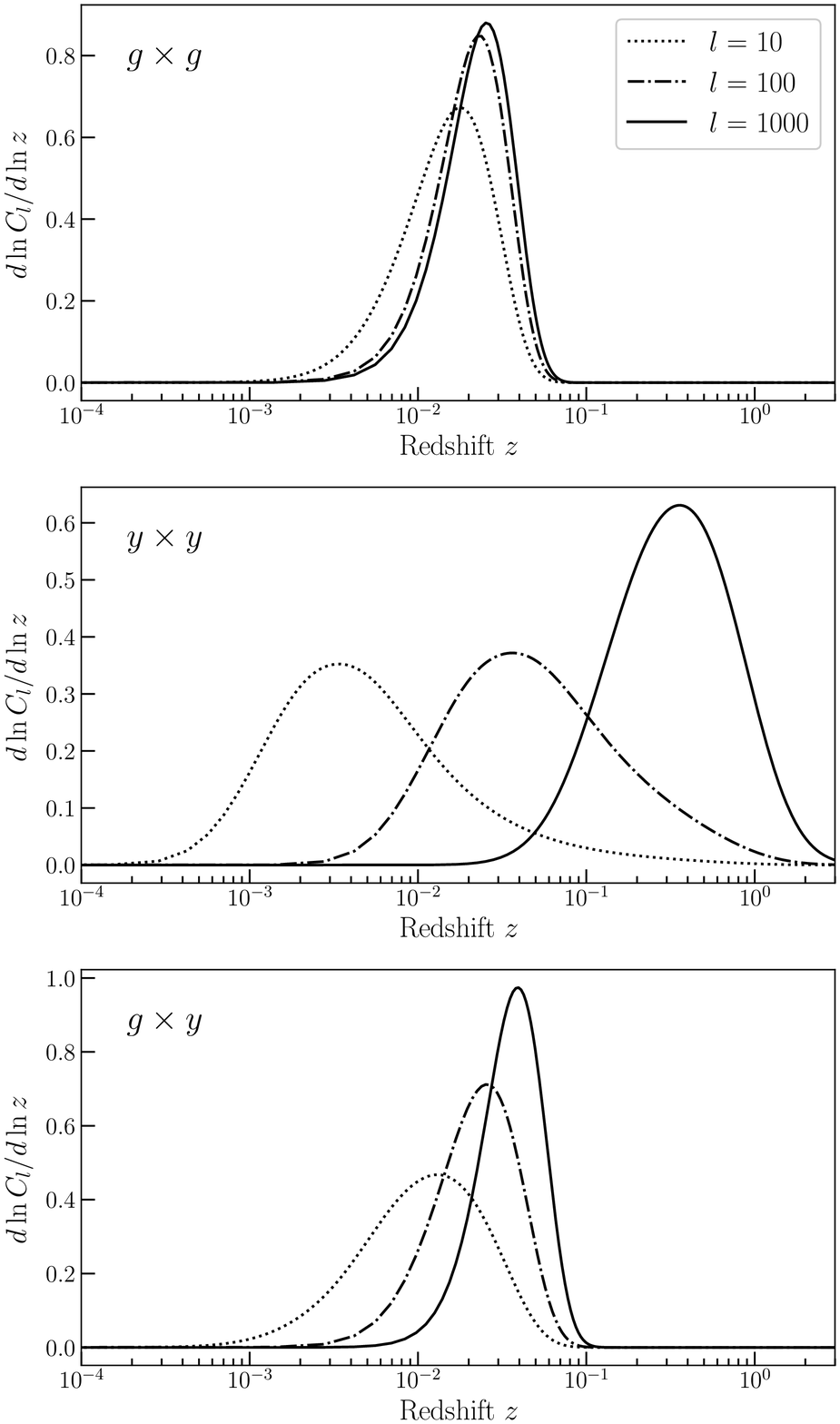}
  \includegraphics[width=\columnwidth]{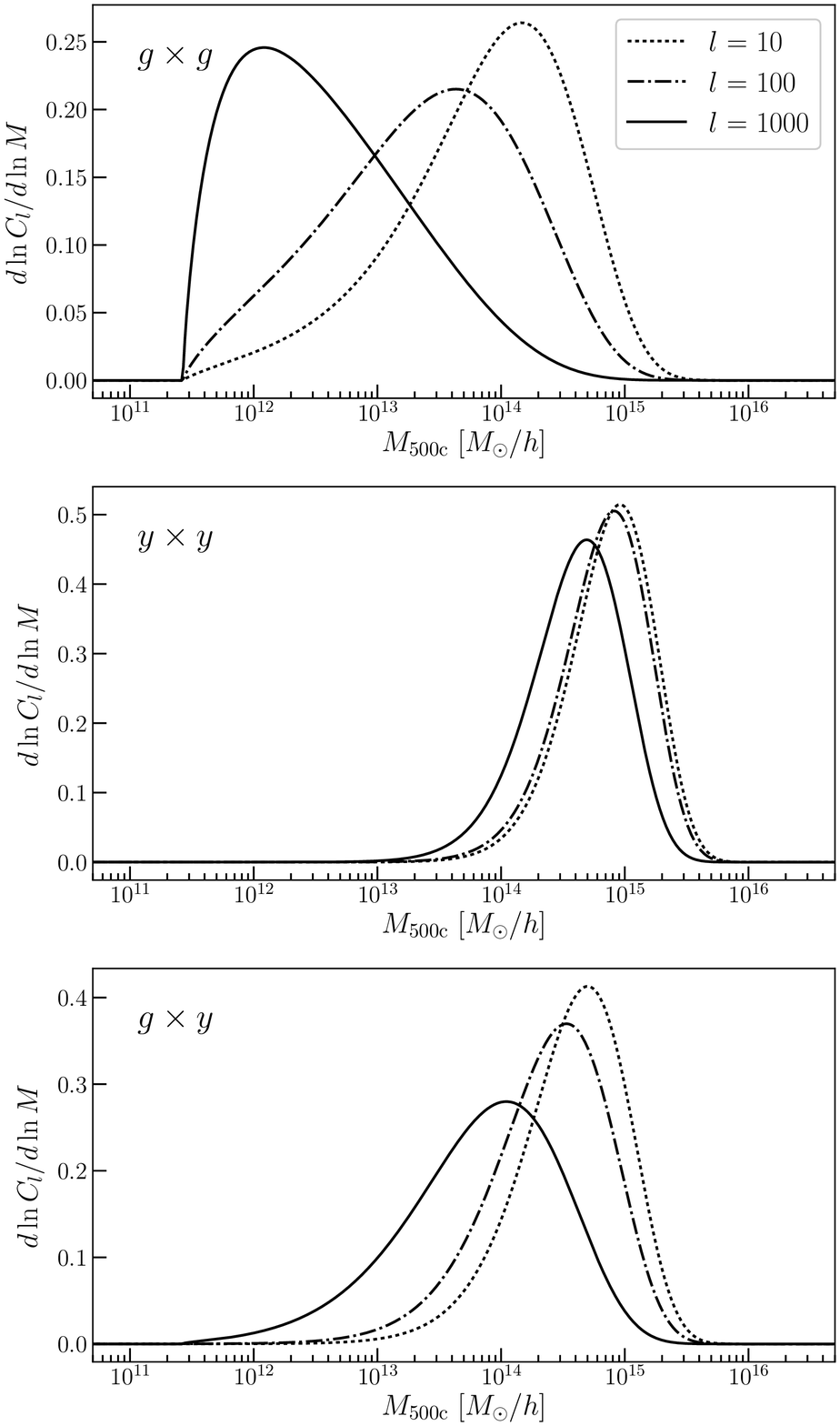}
 \caption{Redshift (Left) and mass (Right) distributions of the 2MRS auto- (Top), the tSZ auto- (Middle) and the 2MRS-tSZ cross power spectrum (Bottom) at multipoles of $l = $10, 100 and 1000.}
 \label{fig:cldz_cldm}
\end{figure*}

Figure \ref{fig:cldz_cldm} shows the redshift distribution of $C_{l}^{\rm AB, 1h}$,
\begin{equation}
\frac{\mr{d}\ln C_{l}^{\rm AB,1h}}{\mr{d}\ln z}
= \frac{z\frac{\mr{d}V}{\mr{d}z \mr{d}\Omega} \int \mr{d}M \frac{\mr{d}n}{\mr{d}M} \tilde{u}_{l}^{\rm A}(M,z)\tilde{u}_{l}^{\rm B}(M,z)}{\int \mr{d}z\frac{\mr{d}V}{\mr{d}z \mr{d}\Omega} \int \mr{d}M \frac{\mr{d}n}{\mr{d}M} \tilde{u}_{l}^{\rm A}(M,z)\tilde{u}_{l}^{\rm B}(M,z)},
\end{equation}
and the mass distribution of $C_{l}^{\rm AB, 1h}$,
\begin{equation}
\frac{\mr{d}\ln C_{l}^{\rm AB, 1h}}{\mr{d}\ln M}
= \frac{M \int \mr{d}z \frac{\mr{d}V}{\mr{d}z \mr{d}\Omega}  \frac{\mr{d}n}{\mr{d}M} \tilde{u}_{l}^{\rm A}(M,z)\tilde{u}_{l}^{\rm B}(M,z)}{\int \mr{d}z\frac{\mr{d}V}{\mr{d}z \mr{d}\Omega} \int \mr{d}M \frac{\mr{d}n}{\mr{d}M} \tilde{u}_{l}^{\rm A}(M,z)\tilde{u}_{l}^{\rm B}(M,z)}.
\end{equation}
Here we fix the model parameters at the best-fitting value (see Section \ref{sec:mcmc} and Table \ref{tb:params}).

The tSZ auto power spectrum has a broad redshift distribution while the 2MRS auto and the 2MRS-tSZ cross have a narrow distribution biased to low-$z$, by construction.

A large fraction of the tSZ auto power spectrum is explained by massive halos ($M_{\rm 500c} > 10^{14} M_{\odot}$) since $\tilde{u}_l^{\rm y}$ scales with $M_{\rm 500c}^{5/3}$.
On the other hand, less massive halos dominate the galaxy-tSZ cross power particularly at smaller scales.

\subsection{Parameter dependence}
In this section, we show how the computed power spectra vary with the model parameters to understand what we can learn from the joint analysis.
Since the 2-halo terms have only minor contributions to the total power spectra within the multipole range we explored, we only show the 1-halo terms.
\subsubsection{Cosmological parameters}
\cite{bolliet/etal:2018} reported that the cosmological parameter dependence of the power spectrum amplitude of tSZ is well approximated by
\begin{equation}
C^{\rm yy}_l \propto \sigma_8^{8.1} \Omega_{\rm m}^{3.2} h^{-1.7} \;\;{\rm for}\;\; l \lesssim 10^3. 
\end{equation}

We find that the scaling of the tSZ-2MRS cross-power spectrum is well approximated by a square root of this scaling relation of the tSZ auto power spectrum.
The galaxy term of the cross-power spectrum is less sensitive to the amplitude of the matter density fluctuation.
Therefore the scaling of the cross-power spectrum amplitude with the cosmological parameters is determined approximately by the tSZ term alone.

\subsubsection{Gas physics parameters}
\begin{figure*}
 \includegraphics[width=2\columnwidth]{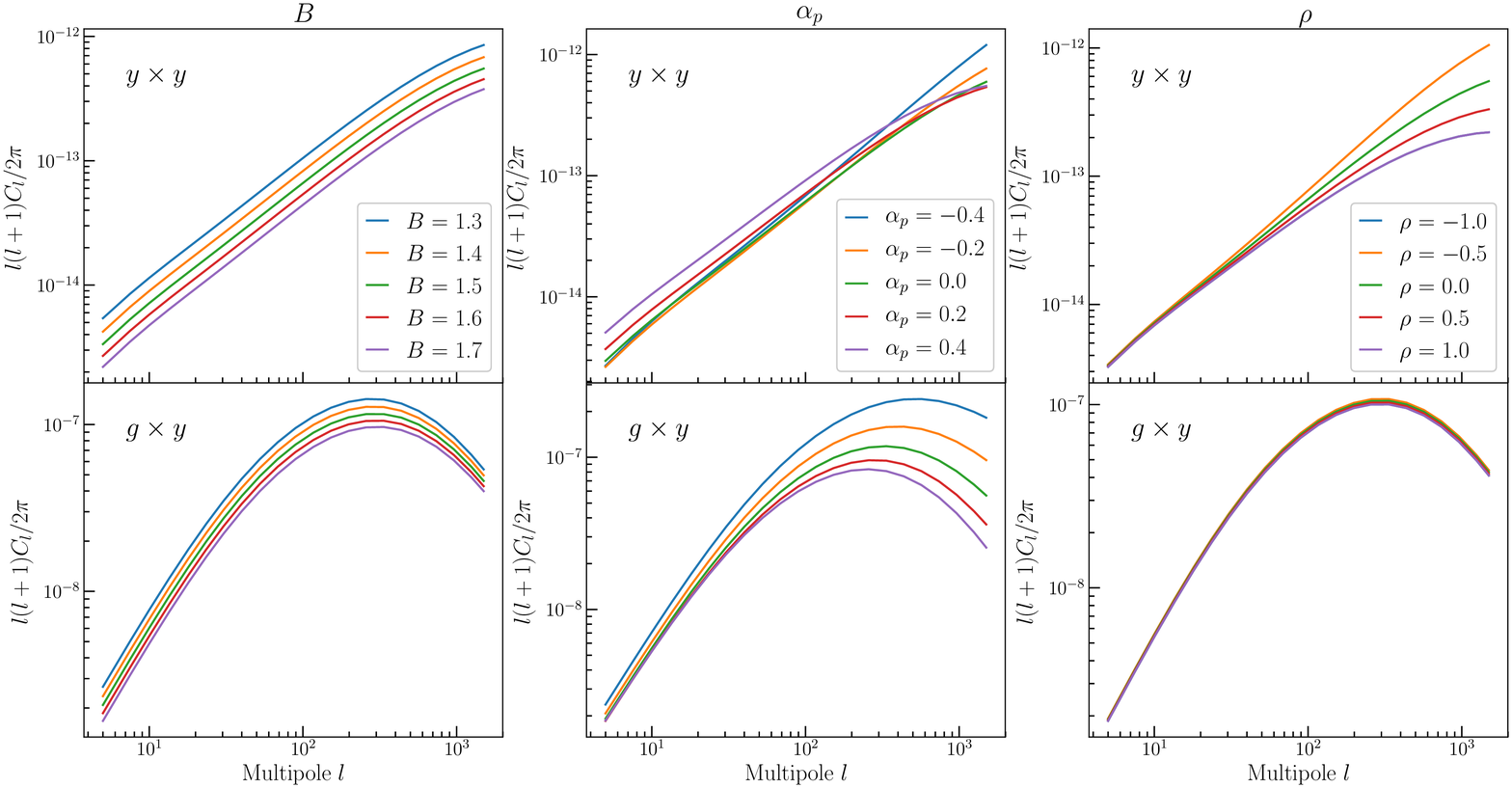}
 \caption{
 Dependence of the tSZ auto (Top) and the 2MRS-tSZ cross power spectrum (Bottom) on $B$, $\alpha_p$ and $\rho$. We only vary a single parameter while the others are fixed at the best-fitting values. 
 Only the 1-halo term is shown.
 }
 \label{fig:cl_b_alpp}
\end{figure*}

Figure \ref{fig:cl_b_alpp} shows the dependence on the parameters related to the halo mass-electron pressure relation given in Eq.(\ref{eq:pe}), $B$, $\alpha_p$ and $\rho$.
The dependence on the bias $B$ is approximated as $C_{l}^{\rm yy} \propto B^{-3.2}$ and $C_{l}^{\rm gy} \propto B^{-1.6}$, which is the same as $\Omega_{\rm m}$. 
The parameters $\sigma_8$, $\Omega_{\rm m}$, $h$ and $B$ affect mostly the overall amplitude of $C_{l}^{\rm yy}$ and $C_{l}^{\rm gy}$ but less the shape of the power spectra. 
Therefore these three parameters would be degenerate.

The dependence on the parameter $\alpha_p$, which determines the slope of the halo mass--pressure relation, differs between the tSZ auto and galaxy-tSZ cross.

As shown in the Figure \ref{fig:cldz_cldm} the mass range of the halos which dominate the tSZ auto is relatively narrow, thus it is not sensitive to $\alpha_{p}$.
At lower multipoles, the cross power spectrum is dominated by the halos whose mass is close to the pivot mass in Eq.(\ref{eq:pe}), $3 \times 10^{14} (0.7/h) M_{\odot}$; thus, the effect of $\alpha_p$ is not significant.
On the other hand, as the cross power is dominated by the less massive halos at higher multipoles, the amplitude decreases with the increase of $\alpha_p$.

The redshift evolution parameter $\rho$ affects significantly on 
the tSZ auto-spectrum at smaller scales, since the power spectrum is dominated by the high-$z$ clusters which appear smaller on the sky.
The $C_{l}^{\rm gy}$ is nearly independent of $\rho$ by construction.

\section{Interpretation}
\label{sec:interpretation}
\subsection{Covariance matrix of the power spectra}
\label{sec:cov}
To perform a likelihood analysis, we need to calculate a covariance matrix of power spectra.
We approximate a covariance as a sum of Gaussian and non-Gaussian components,
\begin{equation}
\label{eq:cov}
{\rm Cov}(C_{l_1}^{\rm AB},C_{l_2}^{\rm CD}) = 
{\rm Cov}^{\rm G}(C_{l_1}^{\rm AB},C_{l_2}^{\rm CD})+
{\rm Cov}^{\rm NG}(C_{l_1}^{\rm AB},C_{l_2}^{\rm CD}).
\end{equation}

The Gaussian term of the covariance matrix is written as (\citealt{hu/jain:2004})
\begin{equation}
{\rm Cov}^{\rm G}(C_{l_1}^{\rm AB},C_{l_2}^{\rm CD})
= \frac{\delta_{l_1 l_2}}{f_{\rm sky}^{\rm AB, CD}(2l_1+1)\Delta l_1}
\left[
\hat{C}_{l_1}^{\rm AC}\hat{C}_{l_2}^{\rm BD}+
\hat{C}_{l_1}^{\rm AD}\hat{C}_{l_2}^{\rm BC}
\right],
\end{equation}
where $\delta$ is the Kronecker delta, $\Delta l$ is the size of multipole bin, $f_{\rm sky}^{\rm AB,CD}$ is the available sky fraction, and $\hat{C}_{l}^{\rm AB}$ is the measured power spectrum including the noise.

The non-Gaussian term can be approximated by the Poisson term of the trispectrum $T_{l_1 l_2}^{\rm ABCD}$ as (e.g., \citealt{komatsu/seljak:2002})
\begin{equation}
\label{eq:cov_ng}
{\rm Cov}^{\rm NG}(C_{l_1}^{\rm AB},C_{l_2}^{\rm CD})
= \frac{1}{4 \pi f_{\rm sky}^{\rm AB, CD}} T_{l_1 l_2}^{\rm ABCD},
\end{equation}
and
\begin{equation}
T_{l_1 l_2}^{\rm ABCD} =
\int \mr{d}z\frac{\mr{d}V}{\mr{d}z \mr{d}\Omega} 
\int \mr{d}M \frac{\mr{d}n}{\mr{d}M}
\tilde{u}_{l_1}^{\rm A}\tilde{u}_{l_1}^{\rm B}
\tilde{u}_{l_2}^{\rm C}\tilde{u}_{l_2}^{\rm D}.
\end{equation}

Since we apply different masks to the 2MRS map and tSZ map, 
$f_{\rm sky}$ differs depending on the combinations of observables.
As already mentioned in Section \ref{sec:data}, $f_{\rm sky}$ becomes $f_{\rm sky}^{\rm gg} = 0.877$ for the 2MRS auto and $f_{\rm sky}^{\rm yy} = 0.494$ for the tSZ auto.
We approximate a $f_{\rm sky}$ of the cross power spectrum as $f_{\rm sky}^{\rm gy} = \sqrt{f_{\rm sky}^{\rm gg} f_{\rm sky}^{\rm yy}}$ following \cite{page/etal:2007}.
We also assume that $f_{\rm sky}$ of the cross covariance is written in the same way, i.e., $f_{\rm sky}^{\rm gg, gy} = \sqrt{f_{\rm sky}^{\rm gg} f_{\rm sky}^{\rm gy}}$.

\begin{figure*}
\begin{center}
 \includegraphics[width=2\columnwidth]{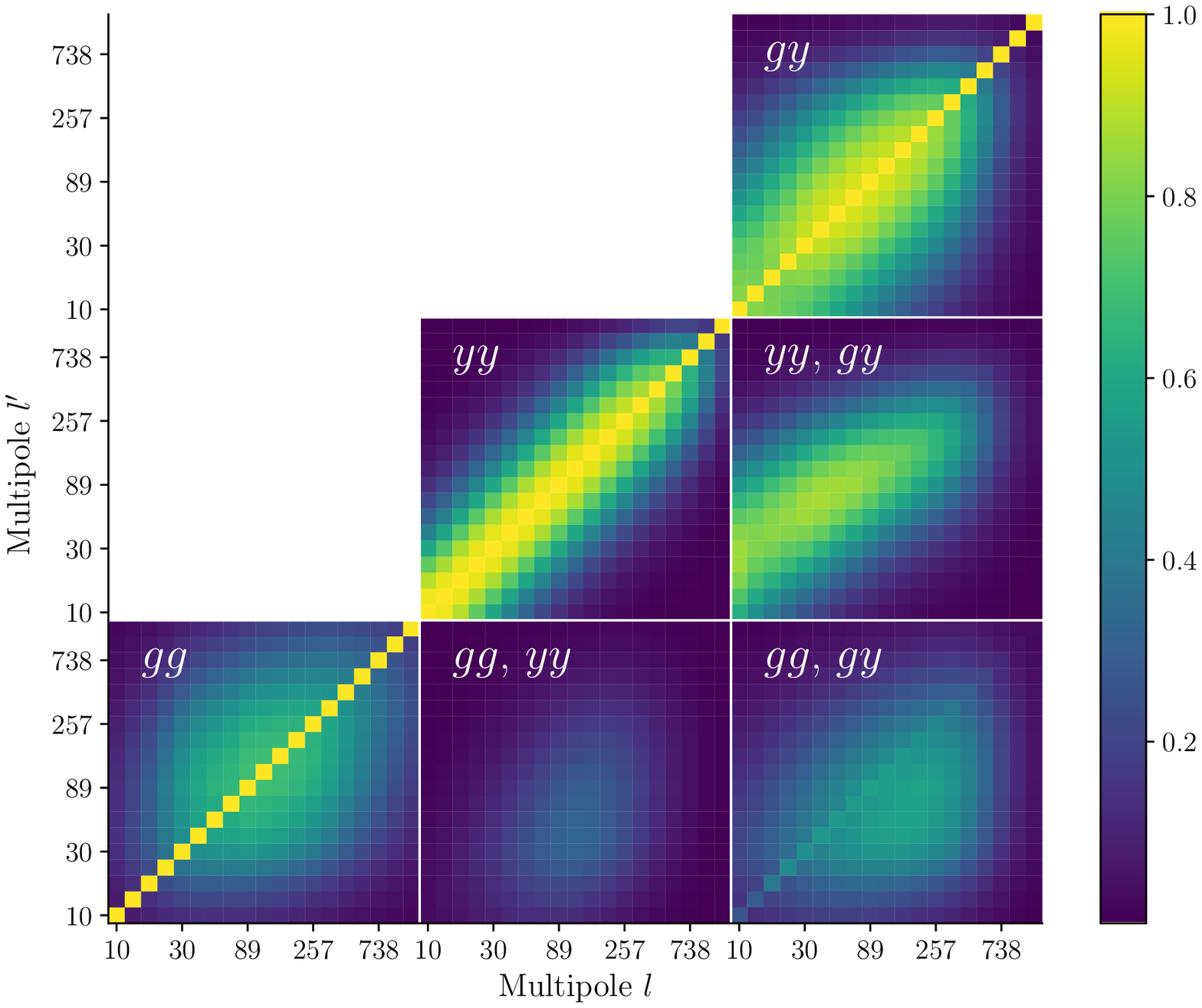}
 \caption{Cross-correlation coefficient matrix of the auto and cross-power spectra, calculated by the analytic formula given in Eq.(\ref{eq:cov}).
 Non-Gaussian term is also included.}
 \label{fig:coeff}
\end{center}
\end{figure*}

The estimated cross-correlation coefficient matrix is shown in Figure \ref{fig:coeff}.
As already pointed out by the previous work (e.g, \citealt{komatsu/seljak:2002, bolliet/etal:2018}), different multipole bins of the tSZ power spectrum are strongly correlated. The covariance of the 2MRS auto and the 2MRS--tSZ cross also shows strong mode coupling. It may be due to nearby massive clusters, which add a power at all multipole ranges (see \citealt{ando/etal:2018} for more details).

We also estimate the covariance matrix by using the Jackknife technique and obtain roughly consistent results with the covariance presented above.
See Appendix \ref{sec:jk} for more details.

\subsection{Parameter fitting}
\label{sec:mcmc}
To compute the posterior probability distribution of the model parameters $\bvtheta$ given the data set $\bd$, $P(\bvtheta|\bd)$, we use Bayes' theorem:
\begin{equation}
P(\bvtheta|\bd) \propto P(\bvtheta)\cL(\bd|\bvtheta),
\end{equation}
where $\cL(\bd|\bvtheta)$ is the likelihood of the data given a model with $\bvtheta$.
For the cosmological parameters, we consider a multivariate Gaussian prior taking into account the constraints from the {\it Planck} CMB observations, as described in detail below. 
For the other parameters, we assume flat priors, i.e., $P(\bvtheta) = $ constant within the range presented in Table \ref{tb:params} and $P(\bvtheta) = 0$ otherwise.
In order to explore the parameter space efficiently, we use the Markov-Chain Monte Carlo (MCMC) technique.
To this end we use the {\tt CosmoMC} package \citep{cosmomc}.

There are several free parameters in our model.
For the galaxy term, the HOD parameters $M_{0}$, $M_1$ and $\alpha_{\rm g}$, and the parameters for the radial distribution of satellite galaxies, $r_{\rm s, g}$ and $r_{\rm max,g}$, are treated as free.
For the tSZ term, the mass bias $B$, its redshift evolution parameter $\rho$, and the index of the mass-pressure relation $\alpha_p$ are treated as free. 
In addition to them, the amplitude of the contaminating sources, $A_{\rm CIB}$, $A_{\rm IR}$ and $A_{\rm RS}$, are also free.
For the cosmological parameters, we treat all 6-parameters, i.e., $\Omega_b h^2$, $\Omega_c h^2$, $\theta_s$, $A_s$, $n_s$ and $\tau_{\rm reio}$ as free.

For the data set $\bd$, we use the measured 2MRS auto-, tSZ auto- and 2MRS-tSZ cross power spectra. 
We use the multipole range of $10 < l < 1247$ for the 2MRS auto and the tSZ auto, but used only up to $l < 500$ for 2MRS-tSZ cross since there would be a contamination from IR and radio point sources and it is difficult to model their correlation with the 2MRS sources.

In addition to the power spectra, we also use the number of 2MRS galaxies within $z < 0.01$, $N_{\rm 2MRS}^{z<0.01} = 3200$, in which the sample is safely volume-limited, to ensure that the model reproduces the number count of galaxies (\citealt{ando/etal:2018}).

We approximate the data likelihood as a multivariate Gaussian:
\begin{eqnarray}
\label{eq:like}
-2 \ln \cL(\bd|\bvtheta) 
&=& {\vDelta}^{\intercal} {\rm Cov}^{-1} \vDelta \nonumber \\
&+& {\vDelta_{\rm cosmo}}^{\intercal} {{\rm Cov}_{\rm CMB}}^{-1} \vDelta_{\rm cosmo} \nonumber \\
&+& \frac{[N_{\rm th}^{z<0.01}(\bvtheta)-N_{\rm 2MRS}^{z<0.01}]^2}{N_{\rm 2MRS}^{z<0.01}} \nonumber \\
&+& \ln{|{\rm det}\; {\rm Cov}|} \nonumber \\
&+& {\it const.},
\end{eqnarray}
where the vector $\vDelta$ denotes the difference between the model power spectra given the parameter set $\bvtheta$ and those measured from the observation.
As the non-Gaussian terms of the covariance matrix depend on the model parameters, we calculate them at each step of MCMC.

The second term in the Eq.(\ref{eq:like}) corresponds to the prior information on the cosmological parameters from the CMB.
The vector $\vDelta_{\rm cosmo}$ is the difference between the proposed value of the cosmological parameters and its mean value from CMB.
We measured the mean value and covariance matrix of the cosmological parameters from the MCMC chains of "TT+lowP+lensing" model which is provided by the {\it Planck} (\citealt{planck2015_cosmo/etal:2016}). 
The mean values are: $\Omega_b h^2 = 0.0223$, $\Omega_c h^2 = 0.1186$, $100 \theta_s = 1.04103$, $\tau_{\rm reio} = 0.0659$, $\ln(10^{10} A_s) = 3.0623$ and $n_s = 0.9677$.
The covariance matrix ${\rm Cov}_{\rm CMB}$ is shown in Table \ref{tb:cmb}.

For all the fitting results presented in the following sections, 
we use the prior information from the {\it Planck} CMB + CMB lensing and the number of 2MRS galaxies at $z < 0.01$.
 
\begin{table*}
\begin{center}
\begin{tabular}{ccccccc}
\hline
\hline
 & $\Omega_{b}h^2$ & $\Omega_{c}h^2$ & $100 \theta_s$ & $\tau_{\rm reio}$ & $\ln({10^{10} A_s})$ & $n_s$  \\
\hline
$\Omega_{b}h^2$ & $5.353 \times 10^{-8}$ & $-2.679 \times 10^{-7}$ & $5.121 \times 10^{-8}$ & $1.945 \times 10^{-6}$ & $3.443 \times 10^{-6}$ & $8.213 \times 10^{-7}$ \\
$\Omega_{c}h^2$ &   & $3.937 \times 10^{-6}$ & $-3.991 \times 10^{-7}$ & $-2.432 \times 10^{-5}$ & $-4.032 \times 10^{-5}$ & $-9.810 \times 10^{-6}$ \\
$100 \theta_s$ &   &   & $2.104 \times 10^{-7}$ & $3.627 \times 10^{-6}$ & $6.629 \times 10^{-6}$ & $1.302 \times 10^{-6}$ \\
$\tau_{\rm reio}$ &   &   &   & $2.684 \times 10^{-4}$ & $4.725 \times 10^{-4}$ & $7.052 \times 10^{-5}$ \\ 
$\ln({10^{10} A_s})$ &   &   &   &   & $8.625 \times 10^{-4}$ & $1.184 \times 10^{-4}$ \\
$n_s$ &   &   &   &   &  & $3.598 \times 10^{-5}$ \\
\end{tabular}
\caption{The inverse covariance matrix for the cosmological parameters, extracted from the "TT+lowP+lensing" model which is provided by the {\it Planck} (\citealt{planck2015_cosmo/etal:2016}).}
\label{tb:cmb}
\end{center}
\end{table*}

\subsection{Results}
\subsubsection{Base-line model}
\begin{figure*}
 \includegraphics[width=2\columnwidth]{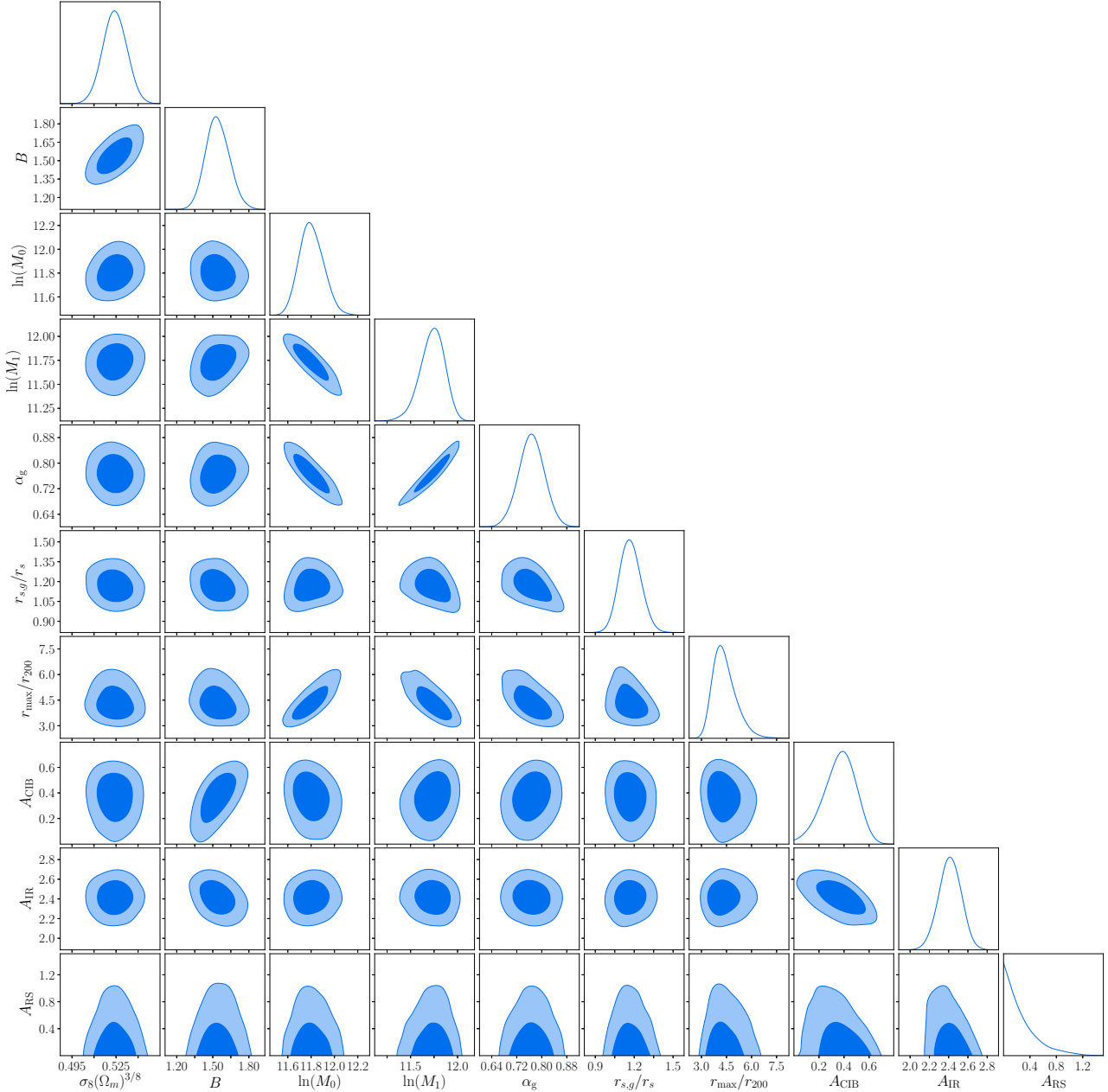}
 \caption{Posterior distribution of the model parameters.
 The diagonal panels show the one-dimensional distribution marginalized over the other parameters. The other panels show the two-dimensional contours of the parameters marginalized over the other parameters. }
 \label{fig:mcmc_all}
\end{figure*}

First, we perform a MCMC fitting fixing $\alpha_p$ at 0.12, which is the value obtained from the analysis of nearby X-ray cluster sample (\citealt{arnaud/etal:2010}), and $\rho = 0$, to explore the simplest model.
The best-fitting model gives $\chi^2 = 54.2$ for 45 degrees of freedom (i.e., 60 data points and 15 parameters).
The probability to exceed is 0.16.
The posterior distributions of the model parameters are shown in Figure \ref{fig:mcmc_all} and summarized in Table \ref{tb:params}. 
We find reasonable values of the HOD and galaxy distribution parameters. See \cite{ando/etal:2018} for the details of the interpretation of the 2MRS auto-power spectrum.

The constraints on the cosmological parameters are not improved from the {\it Planck} results since they are degenerate with the mass bias.
Within the uncertainty of the cosmological parameters, the mass bias $B$ is constrained to be $B = 1.54 \pm 0.098$ (mean and 1$\sigma$), which is slightly tighter than the constraint from the tSZ auto alone, $1.56 \pm 0.103$.
This result suggests that the {\it Planck} cluster mass, which is calibrated against the X-ray observations of local cluster sample (\citealt{arnaud/etal:2010}), should be $35\%$ lower than the true mass to be consistent with the cosmological parameters of {\it Planck} CMB + CMB lensing.
As mentioned in Section \ref{sec:introduction}, the hydrodynamic simulations showed that the non-thermal pressure only accounts for 5-20\% of the mass bias.
Remaining $\sim 15\%$ of the mass bias may be due to other effects such as the calibration error of gas temperature in X-ray observations.
Our best-fitting value of $B$ from the tSZ auto alone is smaller than that of \cite{bolliet/etal:2018}, $B=1.71 \pm 0.17$.
It is because that the {\it Planck} MCMC chain we used includes the constraint from CMB lensing which prefers lower $\sigma_8$ and $\Omega_m$ (\cite{planck2015_cosmo/etal:2016}), while \cite{bolliet/etal:2018} used the CMB temperature fluctuation alone.

\begin{figure}
 \includegraphics[width=\columnwidth]{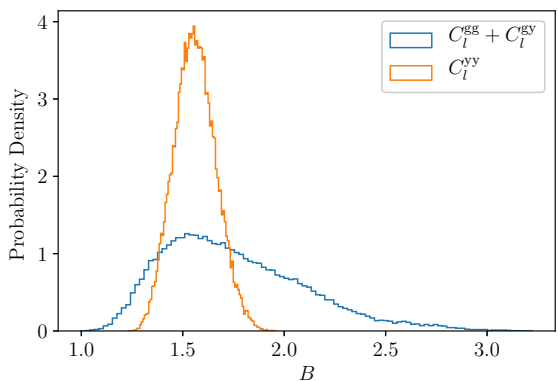}
 \caption{One-dimensional marginalized posterior distributions of the mass bias $B$ from the tSZ auto-power spectrum alone (orange) and the 2MRS-tSZ cross-power spectrum combined with the 2MRS auto (blue).
 The parameters for mass and redshift dependence of $B$, $\alpha_p$ and $\rho$, are fixed at $\alpha_p = 0.12$ and $\rho = 0$.}
 \label{fig:mcmc_gggy_yy}
\end{figure}

To test consistency of the data, we also perform a likelihood analysis by using only the 2MRS-tSZ cross- and the 2MRS auto-power spectra.
The Figure \ref{fig:mcmc_gggy_yy} shows a constraint on the mass bias from these two data sets.
We find $B = 1.75 \pm 0.35$ (mean and 1 $\sigma$) and the best-fitting model gives $B = 1.55$, which are consistent with the results from the tSZ auto-spectrum alone.
The fact that these independent observations agree implies that a large value of $B$ compared to numerical simulations is not due to obvious systematics in the tSZ data.

\begin{table*}
\begin{center}
\begin{tabular}{cccccccc}
\hline
\hline
    &          & \multicolumn{2}{c}{Base-line model} & \multicolumn{2}{c}{$+\alpha_p$} & \multicolumn{2}{c}{$+\rho$} \\
    & Prior    & Mean &  68\% C.L. & Mean & 68\% C.L. & Mean & 68\% C.L. \\
\hline
$B$ & [0,2.5] & 1.54 & [1.44,1.63] & 1.53 & [1.41,1.61] & 1.46 &  [1.31,1.59] \\
$A_{\rm CIB}$ & [0,10] & 0.37 & [0.25,0.51] & 0.35 & [0.23,0.49] &       0.47 & [0.32,0.65] \\
$A_{\rm IR}$ & [0,10] & 2.41 & [2.30,2.53] & 2.35 & [2.21,2.48] & 2.44 & [2.30,2.59] \\
$A_{\rm RS}$ & [0,10] & 0.27 & [0.00,0.32] & 0.26 & [0.00,0.31] & 0.31 & [0.00,0.37] \\
$r_{\rm max}/r_{200}$ & [0,10] & 4.39 & [3.57,4.87] & 4.97 & [3.81,5.81] & 4.60 & [3.65,5.15] \\
 $r_{\rm s,g}/r_{\rm s}$ & [0,10] & 1.17 & [1.08,1.25] & 1.23 &  [1.14,1.39] & 1.19 & [1.10,1.28] \\
$\ln (M_0)$ & [5,15] & 11.80 & [11.69,11.90] & 11.93 & [11.78,12.16] & 11.85 & [11.71,11.95] \\
$\ln (M_1)$ & [5,15] & 11.73 & [11.62,11.87] & 11.48 & [11.13,11.82] & 11.66 & [11.54,11.85] \\
$\alpha_{\rm g}$ & [0,10] & 0.77 & [0.73,0.81] & 0.70 & [0.59,0.78] & 0.75 &  [0.70,0.80] \\
$\alpha_p$ & [-2,2] & -- & -- & 0.03 & [-0.11,0.11] & -- & -- \\
$\rho$ & [-3,3] & -- & -- & -- & -- & 0.69 & [-0.27,1.26] \\
\hline
\end{tabular}
\caption{Best-fitting and 68\% confidence regions of the model parameters.
The range of flat priors are shown in the second column.
For the cosmological parameters we assumed multivariate Gaussian priors shown in Table \ref{tb:cmb}.}
\label{tb:params}
\end{center}
\end{table*}

\subsubsection{Mass tomography}
Next, we treat $\alpha_p$ as a free parameter to investigate the cluster mass dependence of the electron pressure.
The left panel of Figure \ref{fig:mcmc_alpp_rho} shows the constraints of $B$ and $\alpha_p$ marginalized over the other parameters.
When we only use the tSZ auto, $\alpha_p$ strongly correlates with $B$.
Adding the cross power spectrum and the galaxy auto power spectrum lifts degeneracy nicely,
since the mass range traced by the cross-power spectrum is different from the tSZ auto and they have different dependence on $\alpha_p$ as shown in Figure \ref{fig:cl_b_alpp}.
Constraints from the combined auto- and cross-power spectra are $B = 1.53 \pm 0.116$ and $\alpha_p = 0.031 \pm 0.126$ (mean and 1$\sigma$).
This result is consistent with a self-similar model, i.e., $\alpha_p = 0$.

If we assume that the tSZ clusters are in the (nearly) self-similar state, 
the parameter $\alpha_p$ can be translated into the mass dependence of mass bias as $B(M) \propto M_{\rm 500c}^{-1.5 \alpha_p}$. 
Our results suggest that either the mass bias does not strongly depend on the halo mass for self-similar relation, or the departure from self-similar relation ($\alpha_p = 0.12$) cancels the mass dependence of $B$.

\begin{figure*}
 \includegraphics[width=\columnwidth]{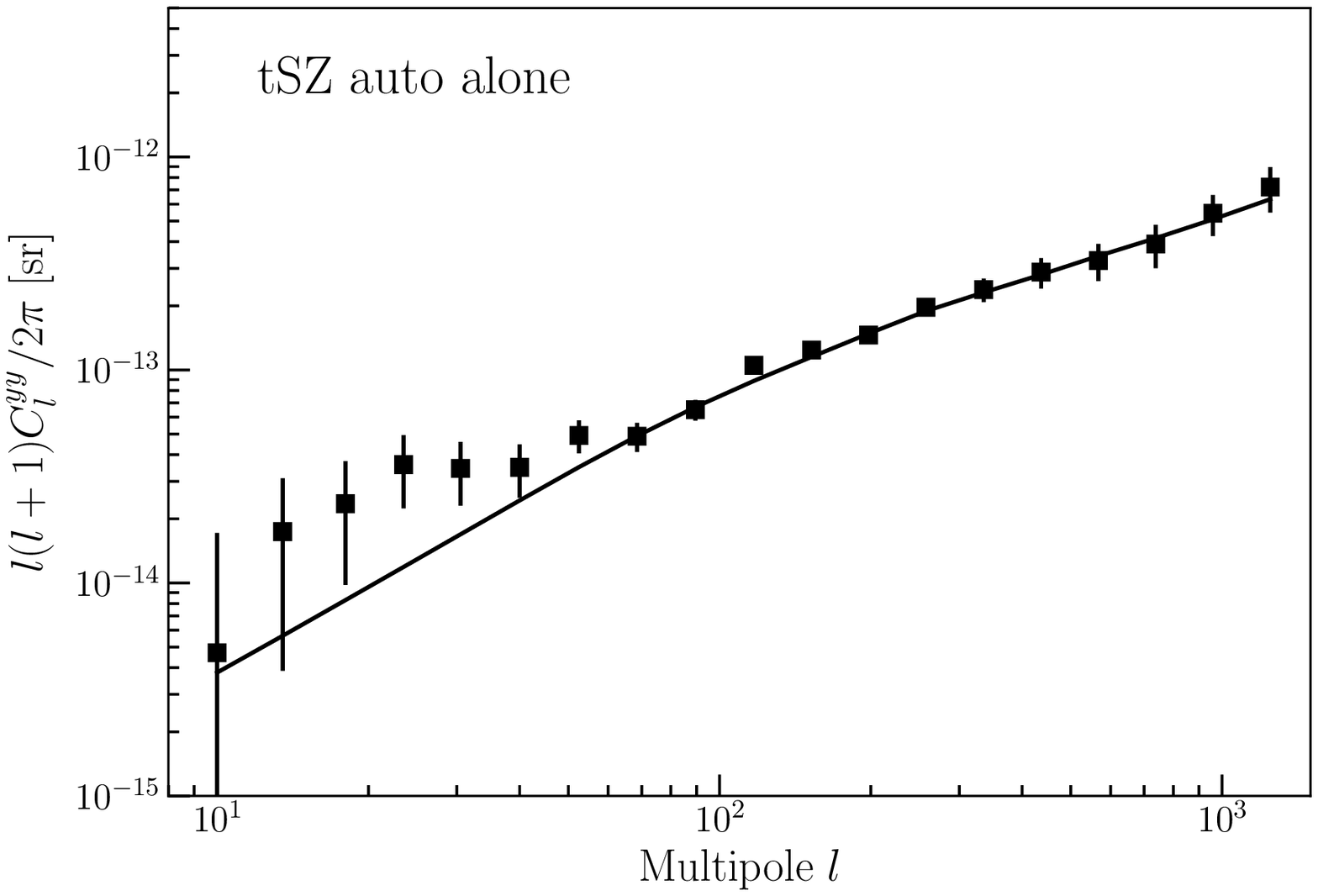}
 \includegraphics[width=\columnwidth]{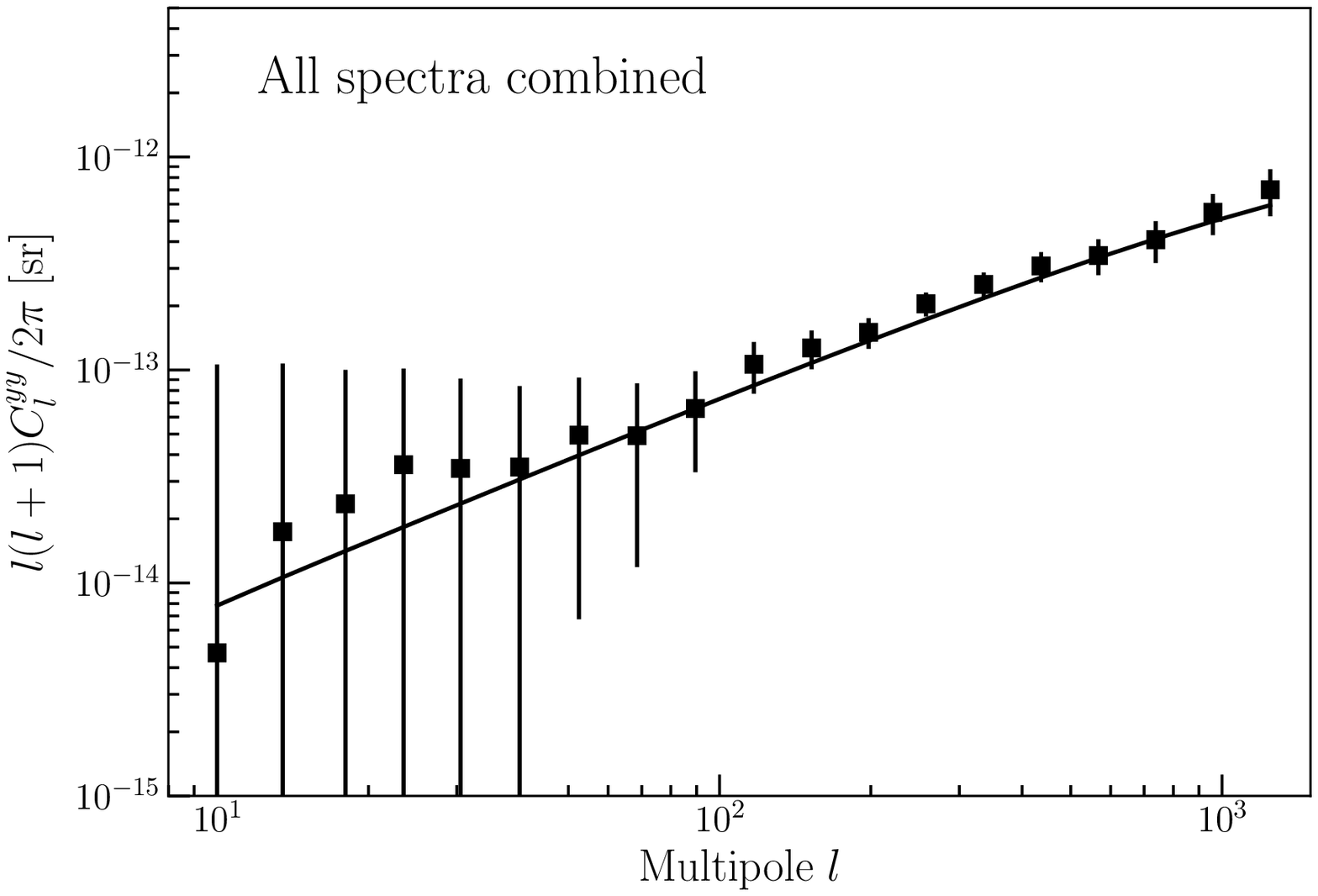}
 \caption{
 The best-fitting tSZ auto-power spectrum determined from the tSZ auto-spectrum alone ({\it Left}) and the joint analysis of all the spectra ({\it Right}), leaving $\alpha_p$ as free.
 The contaminating sources are subtracted. The error bars include the non-Gaussian terms estimated from the best-fitting models given in Eq. (\ref{eq:cov_ng}).}
 \label{fig:cl_alpp_best}
\end{figure*}

Figure \ref{fig:mcmc_alpp_rho} might give an impression that the tSZ auto and the other spectra are in tension.
To address this, in Figure \ref{fig:cl_alpp_best} we compare the best-fitting models from the tSZ auto alone and from the joint analysis of all spectra. 
The best-fitting values are $B = 3.08$ and $\alpha_p = -0.58$ for the tSZ auto alone, while $B = 1.52$ and $\alpha_p = -0.07$ for all the spectra combined.
This difference is due to the non-Gaussian terms in the covariance matrix.
With the largely negative value of $\alpha_p$, the tSZ auto-power spectrum and the non-Gaussian terms of the covariance matrix at lower multipoles are suppressed.
It makes the error bars smaller and increases the chi-square, but at the same time it decreases the log-determinant of the covariance matrix.
The chi-square and the log-determinant of the covariance matrix are $\chi^2 = 38.4$ and $\ln{|{\rm det\;Cov}|} = -1524.0$ for the tSZ auto alone, while $\chi^2 = 28.8$ and $\ln{|{\rm det\;Cov}|} = -1507.7$ for all the spectra combined.
Since the increase of chi-square is smaller than the decrease of log-determinant, largely negative $\alpha_p$ is favored in the tSZ auto alone, though the best-fitting model to all the spectra is a better fit in terms of the value of $\chi^2$.

\subsubsection{Redshift tomography}
Next, we treat $\rho$ as a free parameter to investigate the redshift evolution of the mass bias.

\begin{figure*}
 \includegraphics[width=2\columnwidth]{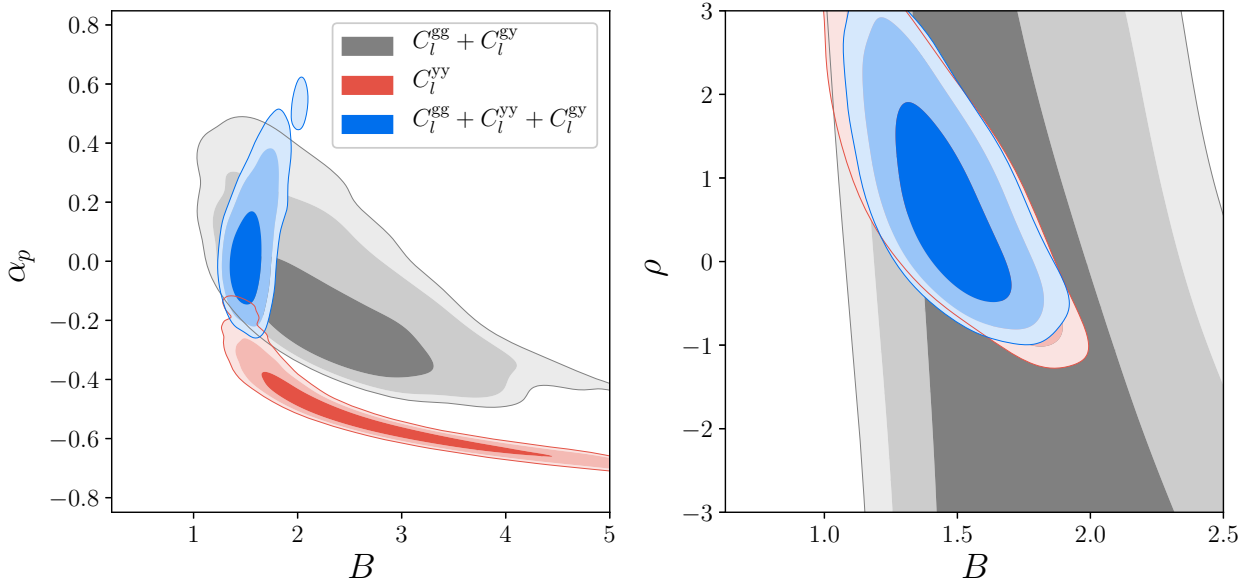}
 \caption{Two-dimensional joint marginalized posterior distribution of the mass bias $B$, and the index of mass pressure relation $\alpha_p$ ({\it Left}) and the redshift evolution parameter $\rho$ ({\it Right}). 
The constraints from the tSZ auto alone (red), the tSZ-2MRS cross combined with the 2MRS (gray), and the all auto- and cross-spectra combined (blue) are shown with the 68\%, 95\%, and 99\% confidence levels.}
 \label{fig:mcmc_alpp_rho}
\end{figure*}

The fitting results are shown in the right panel of Figure \ref{fig:mcmc_alpp_rho}.
We find $B = 1.46 \pm 0.14$ and $\rho = 0.97\pm 0.87 (1\sigma)$ from the tSZ auto alone, and $B = 1.45 \pm 0.14$ and $\rho = 0.69 \pm 0.77$ from the combined analysis of all spectra.

As the $\rho$ affects only high multipoles of the tSZ power spectrum, at which the contaminating sources have significant contributions to the Compton-Y auto-power spectrum, a constraint on $\rho$ is limited by the uncertainties of the amplitude of the contaminations. 
Thus we could not put a tight constraint on $\rho$. 
Once the mass bias at high redshift is determined, the 2MRS-tSZ cross would help to constrain $B$ and $\rho$.

\section{Summary and Conclusions}
\label{sec:summary}

In this paper, we have presented the first measurement of the 2MRS-tSZ cross-correlation power spectrum, as well as results from  a joint likelihood analysis.

We find that the 2MRS-tSZ cross-power spectrum can be explained by the same parameter set that fits the tSZ auto-power spectrum.
From the joint analysis of the 2RMS auto-, tSZ auto- and cross-power spectra, we found that the halo mass of ${\it Planck}$ clusters calibrated by X-ray observations should be 35\% lower than the true mass to reconcile with the cosmological parameters determined from the primordial CMB fluctuation and CMB lensing observed by ${\it Planck}$ (``TT+lowP+lensing''; \citealt{planck2015_cosmo/etal:2016}).
This mass bias is larger than the predictions of numerical simulations, 5--20\%, indicating that the mass bias may not be due only to the assumption of HSE, but also by the other effects such as the calibration error in X-ray observations.
Alternatively, it may have a cosmological implication; namely, the value of $\sigma_8^{8.1}\Omega_m^{3.2}h^{-1.7}$ is actually lower than that inferred from the CMB data and the $\Lambda$CDM model, and may require new physics such as dark energy different from the cosmological constant \citep{bolliet/etal:2018}, massive neutrinos, and so on.

We have also investigated the mass and redshift dependence of the electron pressure profile by introducing free parameters, $\alpha_p$ and $\rho$.
The cross-power spectrum significantly improves the constraint on $\alpha_p$ since the tSZ auto and cross-spectra have different mass distributions and different dependence on $\alpha_{p}$, as shown in Figure \ref{fig:cldz_cldm} and \ref{fig:cl_b_alpp}.
Joint fitting of the auto- and cross-power spectra gives $\alpha_p = 0.031 \pm 0.126$ (mean and 1$\sigma$), which is consistent with the self-similar state ($\alpha_p = 0$).
On the other hand the cross power spectrum does not help constrain  the redshift evolution parameter $\rho$, since the $\rho$ affects only the high multipoles of the tSZ auto-power spectra at which the contaminating sources dominate the signal.
Once the mass bias at high redshift is determined, the cross-power spectra would help to constrain $\rho$ and $B$ at the local Universe.
Constraint from the combined auto- and cross-power spectra is $\rho = 0.69 \pm 0.77$, suggesting that no strong redshift evolution of $B$ is needed.

We thus have demonstrated that the joint analysis of the power spectra of the tSZ and galaxies is a useful tool to study the cluster gas physics.
By taking cross-correlations of the tSZ with various galaxy samples having different mass and redshift range, we can obtain a more detailed picture of the evolution of cluster gas.
The detailed understanding of the gas physics would then eventually enable us to use the tSZ data for the cosmological analysis.

\section*{Acknowledgements}
We thank Boris Bolliet for discussions. This work was supported in part by JSPS KAKENHI Grant Number JP15H05896.

\appendix

\section{Effects of the map reconstruction methods}
\label{sec:milca_nilc}
\begin{figure}
 \includegraphics[width=\columnwidth]{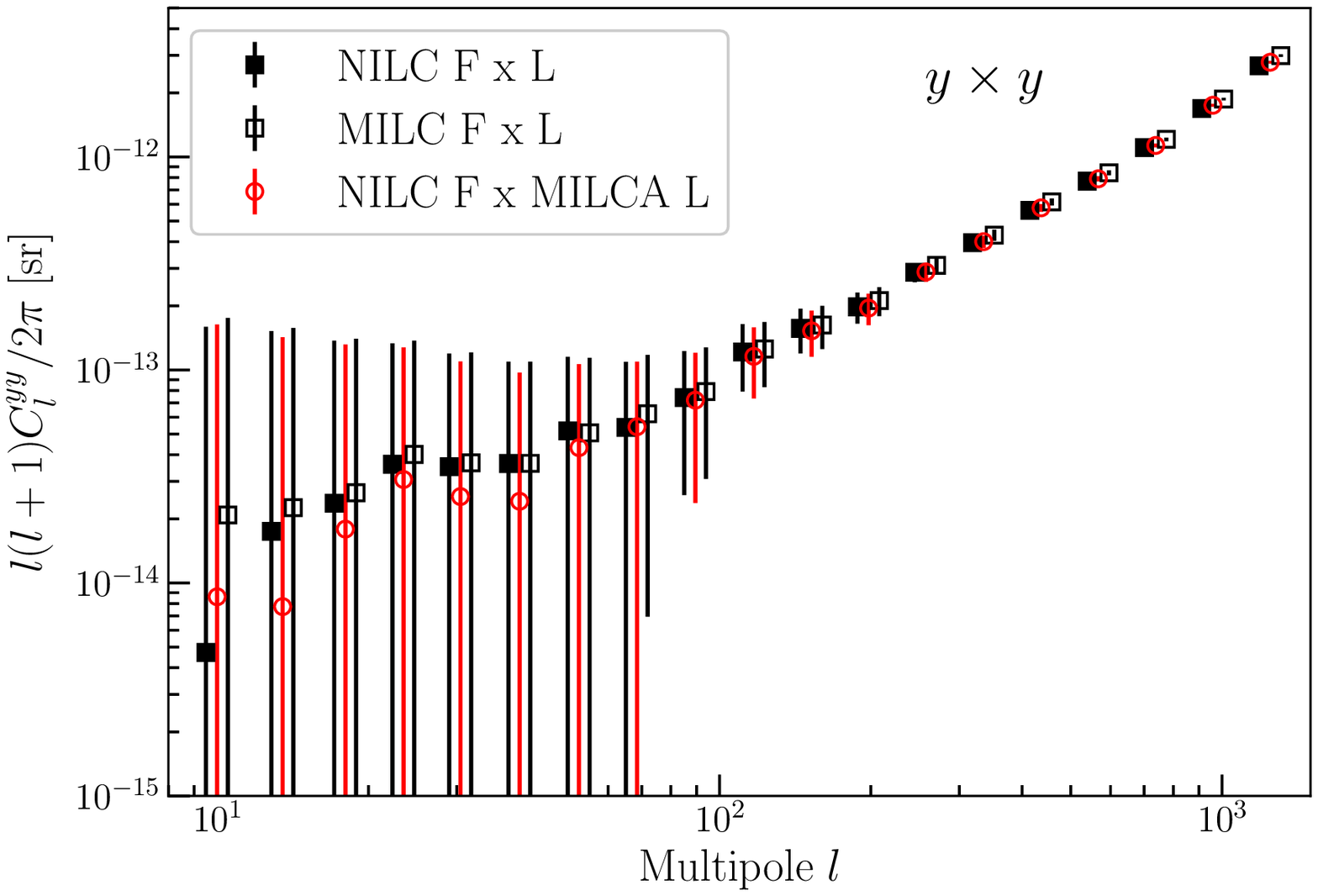}
  \includegraphics[width=\columnwidth]{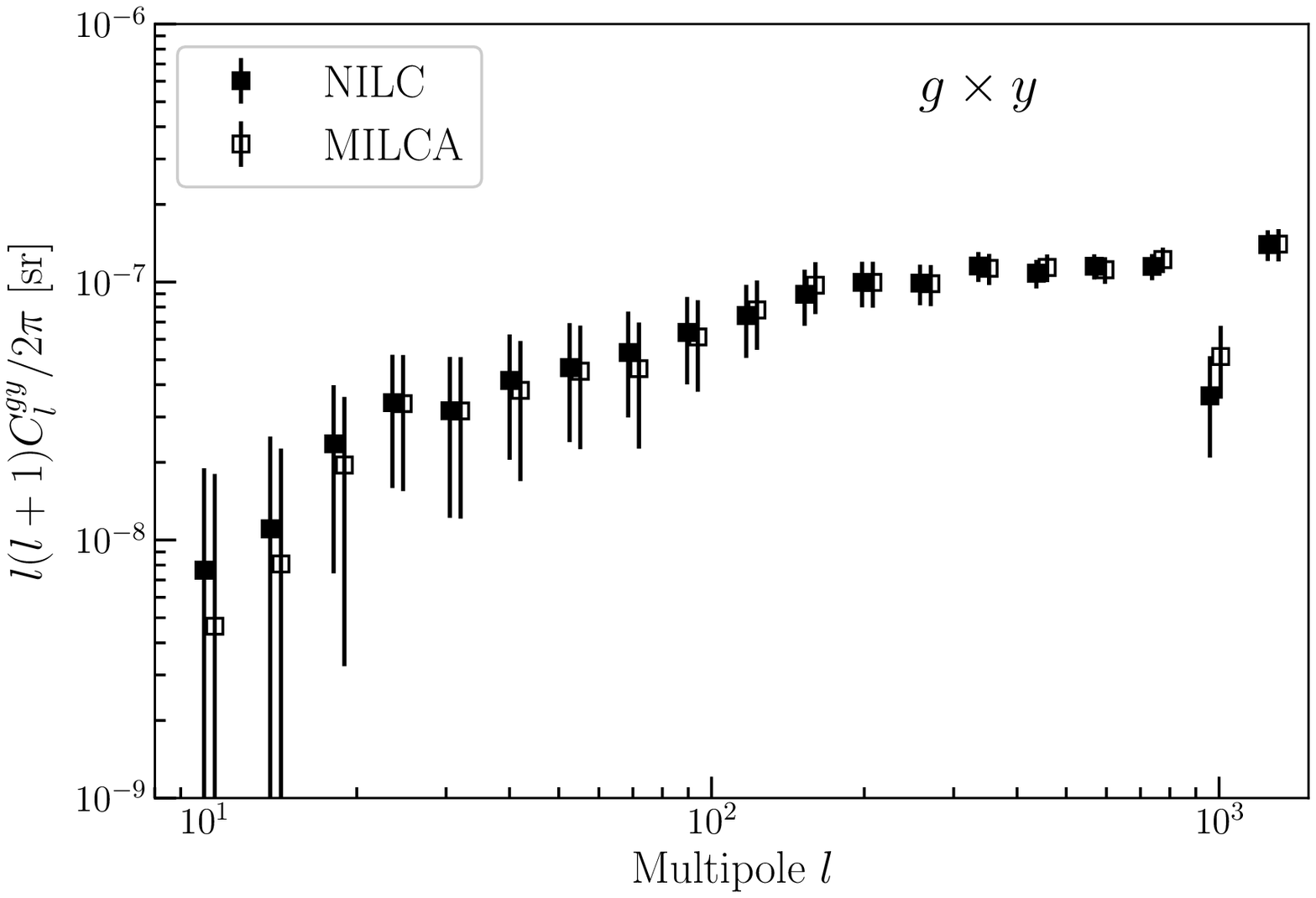}
 \caption{(Top) The tSZ auto-power spectra measured from the NILC and MILCA maps.
The cross-power spectrum of the NILC first and last halves (black filled squares), MILCA first and last halves (black open squares) and NILC first and MILCA last halves (red open circles) are shown.
The contaminating sources are not subtracted.
(Bottom) tSZ-2MRS cross-power spectra measured from the full mission maps of NILC (filled squares) and MILCA (open squares).
The error bars are estimated by the analytic formula as described in Section \ref{sec:cov}.
 }
 \label{fig:nilc_milca}
\end{figure}

The {\it Planck} 2015 data release provide two Compton-Y maps, namely, MILCA and NILC, which use different map reconstruction methods.
Figure \ref{fig:nilc_milca} shows the comparison of the tSZ auto- and tSZ-2MRS cross-spectra estimated from both maps.
For the tSZ auto-spectra, we show the cross-power spectrum of NILC first and last halves, MILCA first and last halves, and NILC first and MILCA last halves.
For the tSZ-2MRS cross-power spectra, we use the full mission maps of MILCA and NILC. 
We find that all the spectra are in agreement within the errors, where the errors are estimated by the analytic formula described in Section \ref{sec:cov}.
For the calculation of the non-Gaussian terms, we use the best-fitting parameters tabulated in Table \ref{tb:params}.
We conclude that the difference in the tSZ map reconstruction method does not affect our results.

\section{Jackknife covariance}
\label{sec:jk}
In Section \ref{sec:interpretation} we used a covariance matrix obtained by the analytic formula, eq.(\ref{eq:cov}).
In this section we compare it with the one obtained by the Jackknife (JK) technique.

In the JK approach, we first divide the sky map into $N_{\rm sub}$ subregions with roughly equal areas in the sky.
Then the auto- and cross-power spectra, $\hat{C}_{l}^{{\rm AB},i}$, are estimated while masking out the $i^{\rm th}$ patch of the sky.
The covariance matrix is estimated from these resampled power spectra as,
\begin{equation}
{\rm Cov}^{\rm JK}(\hat{C}_l^{\rm AB}, \hat{C}_{l'}^{\rm AB}) = \frac{N_{\rm sub}-1}{N_{\rm sub}}\sum_{i=1}^{N_{\rm sub}} (\hat{C}_{l}^{{\rm AB} ,i}-\bar{C}_{l}^{\rm AB})(\hat{C}_{l'}^{{\rm AB} ,i}-\bar{C}_{l'}^{\rm AB}),
\end{equation}
where $\bar{C}_{l}^{\rm AB} = (1/N_{\rm sub}) \sum_{i=1}^{N_{\rm sub}} \hat{C}_{l}^{{\rm AB}, i}$ is the mean of resampled power spectra.

We define the JK subregions by using the coarse HEALPix map. 
In some of these subregions a large fraction of the area is already masked out by the Galactic and point source mask, and therefore the resampled power spectra are almost identical to the one measured by using the whole region.
Thus we decided not to use subregions which are masked larger than $70\%$ of the total area. 
We checked the effect of this threshold and found that it does not strongly affect the results.

\begin{figure}
 \includegraphics[width=\columnwidth]{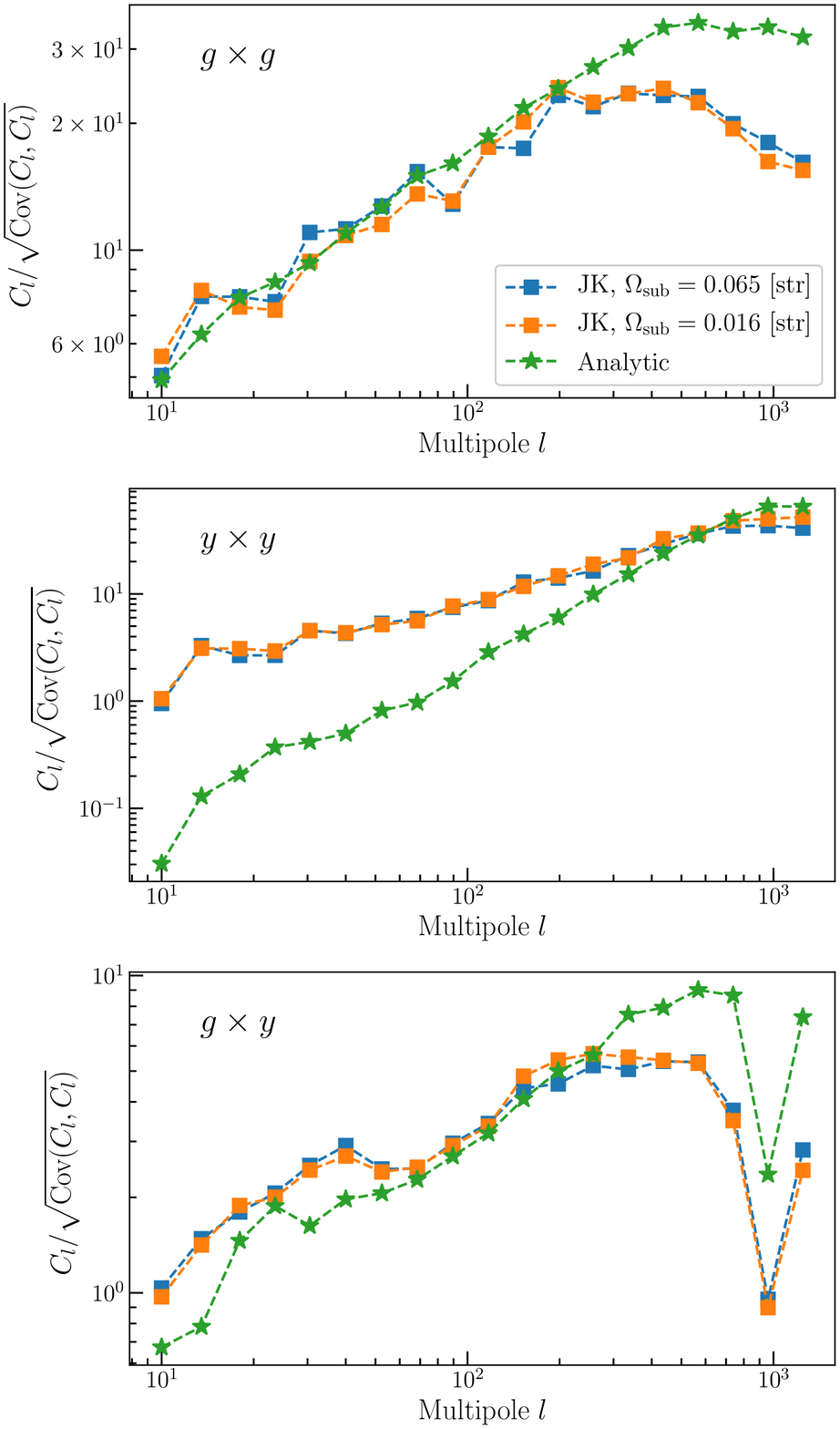}
 \caption{Comparison of measured errors. The square symbols show the signal-to-noise ratio at each multipole with the JK error, while the star symbols show that with the analytic error. }
 \label{fig:err_comparison}
\end{figure}

In Figure \ref{fig:err_comparison} we compare the signal-to-noise ratio at each multipole with the analytic error and the JK error.
For the JK error, we show two results with varying the size of subregions.

For the 2MRS auto-power spectrum, the analytic errors and the JK error are consistent at large scales,
suggesting there is no unknown systematics which is not included in the analytic model.
On the other hand, the analytic errors are larger than the JK errors by a factor of 2--3 at small scales.
It might be due to the halo-exclusion and non-linear effects, which complicates the shot noise (\citealt{ando/etal:2018}).

The analytic error for the tSZ-auto power spectrum is an order of magnitude higher than the JK errors at large scales.
This difference is due to non-Gaussian cosmic variance which is encoded in the tri-spectrum term in eq.(\ref{eq:cov}).
On the other hand, the JK errors are estimated from a single realization and cosmic variance cannot be captured.
The same trend is also seen in the tSZ-galaxy cross-power spectrum but to much lesser extent.

We also performed parameter fitting with the JK covariance matrix, with the same settings described in Section \ref{sec:mcmc}.
For the base-line model (i.e., fixed $\alpha_{p}$ and $\rho$), the mean and 1$\sigma$ error of mass bias is $B = 1.53 \pm 0.07$ for the tSZ auto and $B = 1.51 \pm 0.07$ for the all power spectra, respectively.
They are in agreement with the results with analytic covariance, but with smaller error bars.

\bibliographystyle{mnras}
\bibliography{references} 


\bsp    
\label{lastpage}
\end{document}